\documentclass[useAMS,usenatbib]{mn2e}
\usepackage{graphicx,amssymb, txfonts,pstricks}

\def\pb{Pa$\beta$}
\def\br{Br$\gamma$}

\def\feii{[Fe\,{\sc ii}]}

\def\feh{[Fe\,{\sc ii}]$\lambda 1.64\mu m$}

\def\hm{H$_2$}
\def\hml{H$_2$$\lambda 2.12\mu m$}
\def\p1{Paper~I}

\def\kms {$\rm km\,s^{-1}$}

\title[Nuclear activity and star formation in NGC\,4303]{A SINFONI view of the nuclear activity and circum-nuclear star formation in NGC\,4303}
\author[Riffel et al.]{Rogemar A. Riffel$^{1}$\thanks{E-mail: rogemar@ufsm.br}, L. Colina$^{2,3}$,  T. Storchi-Bergmann$^{4}$, J. Piqueras L\'opez$^{2}$, 
\newauthor S. Arribas$^{2,3}$, R. Riffel$^{4}$, M. Pastoriza$^{4}$, Dinalva A. Sales$^{4}$,  N. Z. Dametto$^{4}$, 
\newauthor A. Labiano$^{5}$, R. I. Davies$^{6}$  \\
$^{1}$ Departamento de F\'\i sica, Centro de Ci\^encias Naturais e Exatas, Universidade Federal de Santa Maria, 97105-900, Santa MMaria, RS, Brazil \\ 
$^{2}$ Centro de Astrobiolog\'ia (CAB, CSIC-INTA), Carretera de Ajalvir, 28850 Torrej\'on de Ardoz, Madrid, Spain \\
$^{3}$ ASTRO-UAM, Universidad Autónoma de Madrid (UAM), Unidad Asociada CSIC, Madrid, Spain\\
$^{4}$ Departamento de Astronomia, Instituto de F\'\i sica, Universidade Federal do Rio Grande do Sul, CP 15051, 91501-970, Porto Alegre, RS, Brazil \\
$^{5}$ Institute for Astronomy, Department of Physics, ETH Zurich, CH-8093 Zurich, Switzerland\\
$^{6}$ Max-Planck-Institut für extraterrestrische Physik, Postfach 1312, D-85741, Garching, Germany
}

\begin{document}

\date{Accepted 1988 December 15. Received 1988 December 14; in original form 1988 October 11}

\pagerange{\pageref{firstpage}--\pageref{lastpage}} \pubyear{2014}

\maketitle

\label{firstpage}

\begin{abstract}

We present new maps of emission-line flux distributions and kinematics in both ionized (traced by H\,{\sc i} and \feii\ lines) and molecular (\hm) gas of the inner 0.7$\times$0.7~kpc$^2$ of the galaxy NGC\,4303, with a spatial resolution 40--80\,pc and velocity resolution 90--150\,\kms obtained from near-IR integral field specroscopy using the VLT instrument SINFONI. The most promiment feature is a 200--250\,pc ring of circum-nuclear star-forming regions. The emission from ionized and molecular gas shows distinct flux distributions: while the strongest H\,{\sc i} and \feii\  emission comes from regions in the west side of the ring (ages $\sim$4\,Myr), the \hm\ emission is strongest at the nucleus and in the east side of the ring (ages $>$10\,Myr). 
We find that regions of enhanced  hot H$_2$ emission are anti-correlated with those of enhanced \feii\ and H\,{\sc i} emission, which can be attributed to post starburst regions that do not have ionizing photons anymore but  still are hot enough ($\approx$\,2000\,K) to excite the H$_2$ molecule. The line ratios are consistent with the presence of an AGN at the nucleus.
 The youngest regions have stellar masses in the range 0.3-1.5$\times10^5$\,M$_\odot$ and ionized and hot molecular gas masses of $\sim0.25-1.2\times 10^{4}$~M$_\odot$ and  $\sim2.5-5~M_\odot$, respectively. The stellar and gas velocity fields show a rotation pattern, with the gas presenting larger velocity amplitudes than the stars, with a deviation observed for the \hm\ along the nuclear bar, where increased velocity dispersion is also observed, possibly associated with non circular motions along the bar.
The stars in the ring show smaller velocity dispersion than the surroundings, that can be attributed to a cooler dynamics due to their recent formation from cool gas.

\end{abstract}


\section{Introduction} \label{intro}

Star formation in the circumnuclear regions of galaxies and its connection with the existence of nuclear young stellar clusters and/or Active Galactic Nuclei (AGN) has been the subject of many studies over the past several decades since early models invoking dynamical resonances in a rotating bar potential \citep{combes85}, bars within bars \citep{sho89}, and the direct feeding of AGN due to stellar winds or cloud-cloud collisions in the vicinity of the nucleus \citep{norman88,sho90}. Some models  \citep[e.g.][]{heller94,knapen95}  suggest that gas could flow inwards from the ring, creating a disk of gas that, if massive enough, would become unstable. Under this scenario, a massive black hole in the nucleus could be fed triggering an AGN \citep{sho89,fukuda98}.

Regarding star formation in nuclear rings, two scenarios have been proposed:  the ``pop-corn'' \citep{elmegreen94} and the ``pearls on a string'' \citep{boker08}.  The ``pop-corn'' scenario assumes that the cold molecular gas is accumulated in a circumnuclear resonance ring. If massive enough, the ring becomes gravitational unstable, fragmenting in clumps and forming stellar clusters at random positions. On the other hand,  in the ``pearls-on-a-string" scenario, new stars are exclusively formed  in the regions where the gas enters the ring (i.e. the regions of maximum gas density). These young clusters evolve passively as they orbit along the ring,  producing a string of aging clusters.  Independent on the formation mechanism of the stellar clusters, the role of these clusters as they move along the ring can be important.  Stellar winds and subsequent supernovae explosions can affect  their surrounding interstellar medium, removing gas and halting subsequent star formation, as well as generating shocks that will  reduce the angular momentum of the gas, that could eventually fall towards the center, feeding the AGN and/or forming a nuclear star cluster. 

Detailed multi-wavelength two-dimensional spectroscopic studies in nearby galaxies with the adequate spatial resolution ($\sim$10 pc, or less) are needed to further investigate the evolutionary scenarios mentioned above. 
CO interferometric maps (PdBI) have shown the presence of a wide range of molecular gas structures in the central  kpc region of nearby galaxies with an AGN \citep{garcia-burillo05}. According to these studies, most of the molecular gas is concentrated in the form of a circumnuclear ring of several hundreds pc to kpc size, while $\approx$33\% of the galaxies show the evidence for a direct gas fueling into the AGN down to scales of $\sim$50 pc \citep{garcia-burillo12}. Of particular importance, the near-infrared (near-IR) bands allow the study of the multi-phase gas, from molecular (H$_2$) to shocked partially-ionized (Fe\,{\sc ii}) to highly ionized gas (e.g. Ca\,{\sc viii}), that can trace a number of different physical structures and mechanisms, from molecular gas reservoirs to AGN outflows.

In the near-IR, recent studies using integral field spectroscopy \citep[IFS,][]{genzel95}, have been particularly useful to map the stellar and gas kinematics, as well the gas excitation and distribution of the different gas phases \citep[e.g.][]{n1068-exc,barbosa14}.
In nearby Seyfert galaxies, compact (scales of few tens of parsecs) molecular gas disks \citep{mazzalay14,mrk1066-kin,hicks09}, streaming motions towards the nucleus \citep[e.g.][]{davies14,diniz15}, ionized gas outflows \citep[e.g.][]{iserlohe13,sb10} and young stellar populations \citep[e.g.][]{davies07} have been mapped in the nuclear regions.  In addition, studies of circumnuclear star-forming rings at hundred of pcs from the nucleus in nearby spirals \citep{boker08,falcon-barroso14,laan13a,laan15} have focused on establishing the reality of the proposed evolutionary scenarios for the star-forming rings, such as the ``pop-corn'' scenario \citep{elmegreen94} and the ``pearls-on-a-string" scenario \citep{boker08}.  

All these previous studies have been focused on either nearby luminous Seyfert galaxies where the output energy is dominated by the AGN, or galaxies with luminous circumnuclear star-forming rings. Here we perform a similar study for NGC\,4303, a nearby galaxy with both a (low-luminosity) AGN and a star-forming ring, and that has also a young massive cluster at the nucleus. In addition, multi-wavelength observations \citep{colina97,colina99,colina02} reveal a spiral of circumnuclear star-forming regions (CNSFRs) that can be traced all the way into 
the inner few parsecs \citep[e.g.][]{colina00,jimenez03}, suggesting it could be the feeding channel to the AGN and nuclear star cluster.  Finally, a high-velocity nuclear outflow extending up to $\approx$120\,pc to the north-east of the nucleus has also been observed in optical line emission of [O\,{\sc iii}] \citep{colina99}.  


NGC\,4303 is at a distance of 16.1~Mpc \citep{colina99}, and is classified as a SB(rs)bc \citep{vaucouleurs91}. It is abundant in cold molecular gas \citep{schinnerer02} that shows the global distribution expected for the gas flow in a strong, large-scale bar, and the two-arm spiral structure in the inner kiloparsec can be explained by a density wave activated by the potential of that bar.


In this paper, we present near-infrared IFS of the inner 350 pc radius of NGC\,4303, obtained with the integral field spectrograph  SINFONI at the VLT in order to map
the kinematics and excitation properties of the different phases of the interstellar medium in the circumnuclear region. It is the first time that such mapping is provided in the near-IR. This work is organized as follows: Section~2 presents the observations and data reduction procedure, while in Sec.~3, we present maps for emission-line flux distributions and ratios, as well as for the gas and stellar kinematics. The results are discussed in Sec.~4 and Sec.~5 presents the conclusions of this work.

\section{Observations and Data Reduction}




The observations were done using the near infrared spectrograph SINFONI of the VLT, during the period 82B (February 2009).
The pointings were centered on the nucleus of the galaxy, covering a field of view (FoV) of $\sim$8"x8" per exposure, enlarged by dithering
up to $\sim$9.25"x9.25", with a plate scale of 0.125x0.220 arcsec pixel$^{-1}$. This corresponds to an average coverage of $\sim$0.7x0.7\,kpc with a spatial sampling of $\sim$10\,pc per spaxel.

The data were taken in the J (1.10-1.35\,$\mu$m), H (1.45-1.80\,$\mu$m), and K (1.97-2.44\,$\mu$m)  bands with a total integration time of 2400\,s per band.   In the same way, a set of photometric standard stars was observed to perform the telluric and flux calibration. We estimated the spatial resolution of our seeing-limited observations by fitting a 2D Gaussian 
profile to a collapsed image of the standard stars. The spatial resolution (FWHM) measured for each band is $\sim$1", $\sim$0.6", and $\sim$0.5" for J-, H-, 
and K-band, respectively, that correspond to 78\,pc, 47\,pc, and 39\,pc at the adopted distance of 16.1 Mpc.


The reduction and calibration processes were performed using the standard ESO pipeline {\sc esorex} (version 3.8.3), and our own IDL routines.
The usual corrections of dark, flat fielding, detector linearity, geometrical distortion, and wavelength calibration were 
applied to each object and sky frames, before subtracting the sky emission from each on-source frame. After each individual data cube was calibrated, we constructed a final data cube per band, taking the relative shifts in the dithering pattern into account.

The calibration of each individual data cube was performed in two steps. Firstly, we removed the telluric absorption from the spectra. We extracted the
integrated spectrum of the corresponding standard star within an aperture of 3$\sigma$ of the best 2D Gaussian fit of a collapsed image. We normalized the
spectrum by a black body profile at the T$_{\rm eff}$ from the Tycho-2 Spectral Type Catalog \citep{Wright03}, after removing the strongest
absorption features of the stars. The result is a \emph{sensitivity function} that accounts for the telluric absorption.

Secondly, each data cube was flux calibrated. The spectrum of the standard star was converted from counts to physical units using the response curves of
the 2MASS filters \citep{Cohen03}, and the J, H, and K magnitudes from the 2MASS catalog \citep{Skrutskie06}. We then obtained a conversion
factor from counts to physical units. Each individual data cube was then divided by its corresponding \emph{sensitivity function} and multiplied by this conversion factor to obtain a full-calibrated data cube. The estimated uncertainty for the conversion factor is $\leq$15\% for each band.

After the data reduction procedure, we performed a spatial filtering of the data cubes using a Butterworth bandpass filter \citep[e.g.][]{gonzalez02,menezes15} in order to remove noise from the observed cubes. The final signal to noise ratio ($snr$) in the continuum is in the range 3--10 for the J band with the smallest values observed close to the borders of the FoV and the highest values at the nucleus, while for the H and K bands we get $snr=3-15$. The improvement of the $snr$ by the spatial filtering ranges from 30 to 5\,\%, from the borders of the FoV to the nucleus for all bands, as obtained by calculating the ratios between the $snr$ of the filtered cubes and those of the original ones. By comparing the continuum images for the original and filtered cubes, we concluded that the filtering procedure does not change the angular resolution of the data.

\section{Results}

\begin{figure*}
\centering
\includegraphics[scale=0.9]{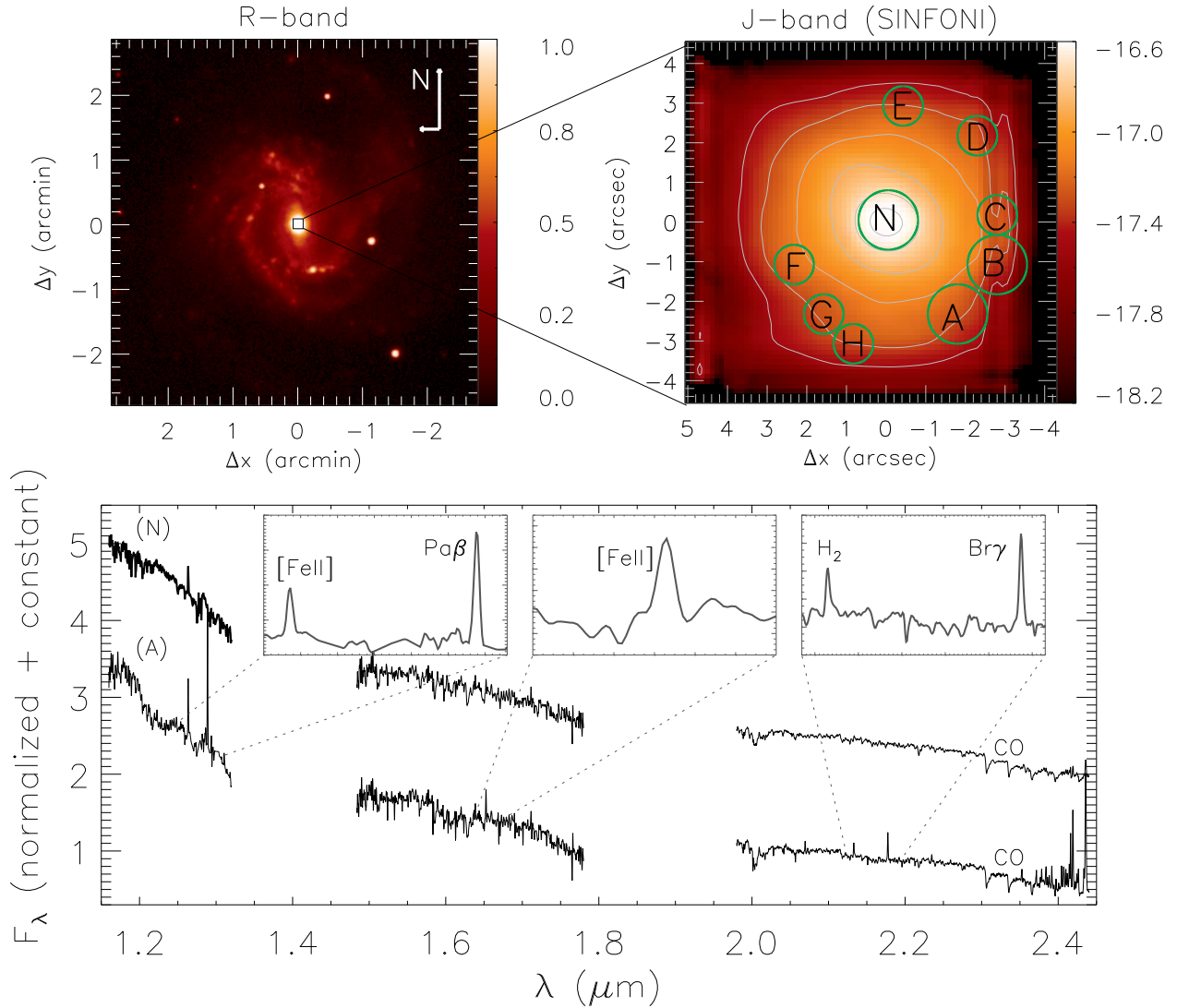}
\caption{Large scale image in the R-band (top-left) from \citet{koopmann01}. J-band continuum, reconstructed from the SINFONI datacube as an average of the fluxes between 1.15 and 1.20 $\mu$m (top right).  The circles mark the position of the circumnuclear star forming regions and the bottom panel shows the near-IR spectra for the nucleus (top) and for position A (bottom), obtained for a circular aperture with radius 0\farcs75 and normalized at 2.1~$\mu$m.  The sub-panels present a zoom of the spectra from position A to better show the main emission lines. The K-band CO absorption band-heads are clearly visible in both spectra. A constant (1.5) was added to the nuclear spectrum for visualization purpose. The color bar for the large scale image is in arbitrary units and for the J-band it shows the fluxes in logarithmic units [log (erg s$^{-1}$ cm$^{-1}$ spaxel$^{-1}$)].}
\label{large}
\end{figure*}

The top-left panel of Figure~\ref{large} shows a large scale optical image of NGC~4303 in the R-band from \citet{koopmann01}\footnote{This image is available at NASA/IPAC Extragalactic Database (http://ned.ipac.caltech.edu)}. In the top-right panel, we show the J-band continuum image obtained from the SINFONI datacube for the inner $9\times9$\,arcsec$^2$, where we label the nucleus `N'  as well as several star-forming regions regions, in particular region `A' (a bright circum-nuclear \br\ emitting clump), for which the spectra are shown at the bottom panels. The spectra were obtained by integrating the fluxes within a circular aperture of radius 0\farcs75 and normalized to the flux at 2.1~$\mu$m. The two spectra show similar slopes, suggesting that the nuclear and extra-nuclear continuum emission have similar origins. On the other hand, the line emission is stronger in region `A' than at the nucleus. In the bottom panel, we present also a zoom of spectrum from position A showing the main emission lines: \feh, \feii$\lambda1.25\mu$m, \pb, \br\ and \hml.

\subsection{Emission-line flux distributions and ratios} 

Different phases of the interstellar medium can be traced in the near-IR by distinct emission lines: ionized (traced by \br\ and \pb\ lines), partially ionized/shocks (\feii\ lines) and hot molecular gas (H$_2$ lines). We used the emission-line PROfile FITting (PROFIT) routine \citep{profit} to fit the line profiles with Gaussians and constructed maps for the flux distributions, centroid velocities and velocity dispersions for each emission line.  

\begin{figure*}
\centering
\includegraphics[scale=1]{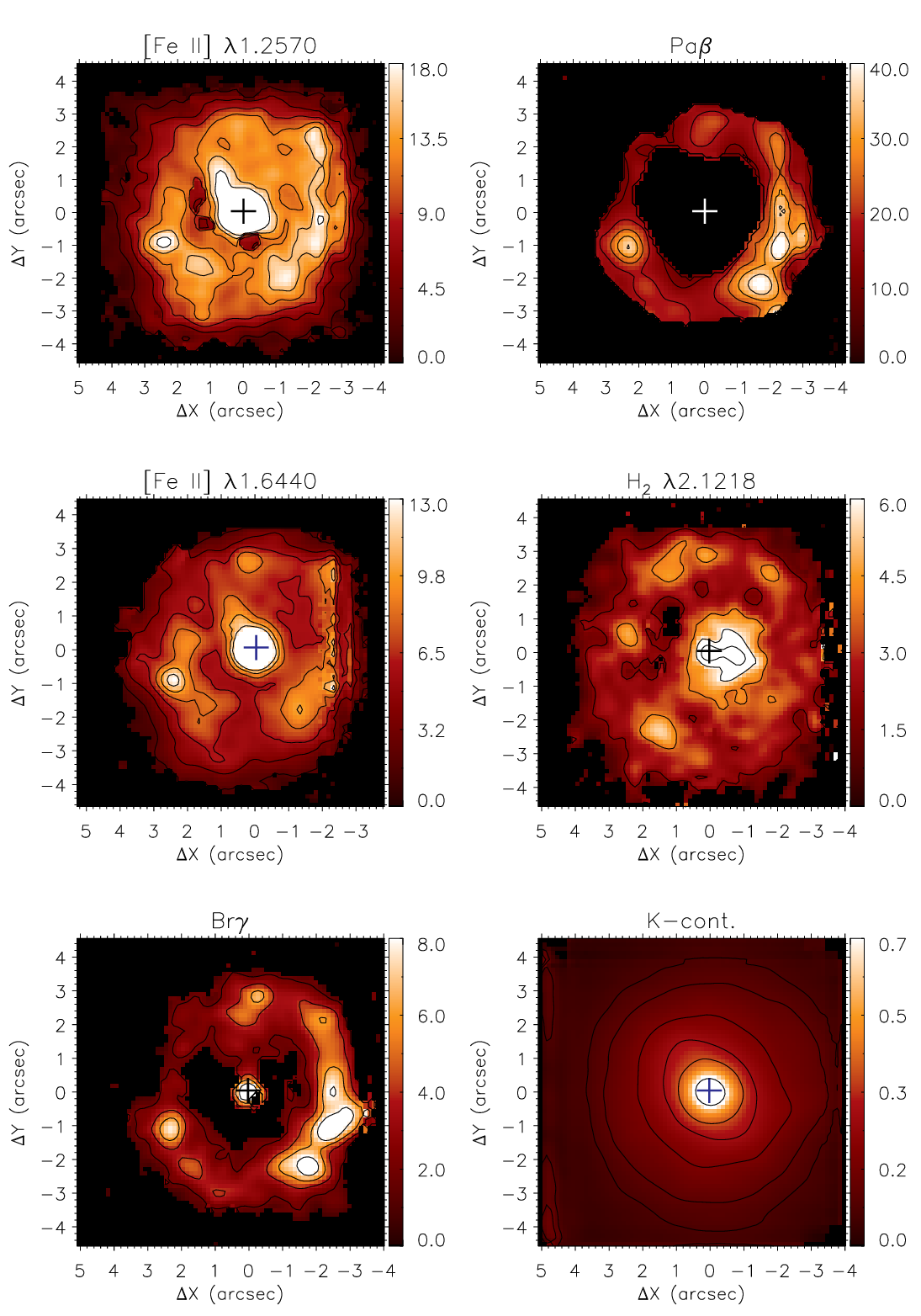}
\caption{Emission-line flux distributions. The color bars show the fluxes in units of $10^{-18}\,{\rm erg s^{-1} cm^{-2}}$ and the central cross marks the position of the nucleus. Black locations are masked regions due to the non-detection of the emission lines or due to bad fits of their profiles.} 
\label{fluxmaps}
\end{figure*}

In Figure~\ref{fluxmaps} we show the flux distributions for the \feii$\lambda1.25\mu$m, \pb, \feh, \hml\ and \br\ emission lines. Black in these maps represent masked locations, where the $snr$ of the lines was not high enough to allow good fits, and regions where the lines were not detected.  We masked regions with uncertainties in flux larger than 50\%, but, for most locations, the uncertainties are smaller than 20\%. The bottom-right panel shows the K-band continuum map. All emission lines present extended emission up to 4$^{\prime\prime}$ ($\approx$ 300\,pc), and the main feature of the flux distributions is a circum-nuclear ring with clumps of enhanced line emission with radius in the range $\approx$2.5--3.2$^{\prime\prime}$ (200--250\,pc). Some differences are observed among the distinct emission lines: (i) while the \feii\ and \hm\ flux distributions present emission at the nucleus, \pb\ is not detected there, while \br\ is marginally detected ($snr\approx2$);
(ii) along the ring, the H\,{\sc i} and [Fe\,II] emission is strongest to the west (including south-west), and at a region to the south-east, while the \hm\ emission is strongest mostly to the east (including south-east and north-east).

\begin{figure*}
\centering
\begin{tabular}{l r}
\includegraphics[scale=0.64]{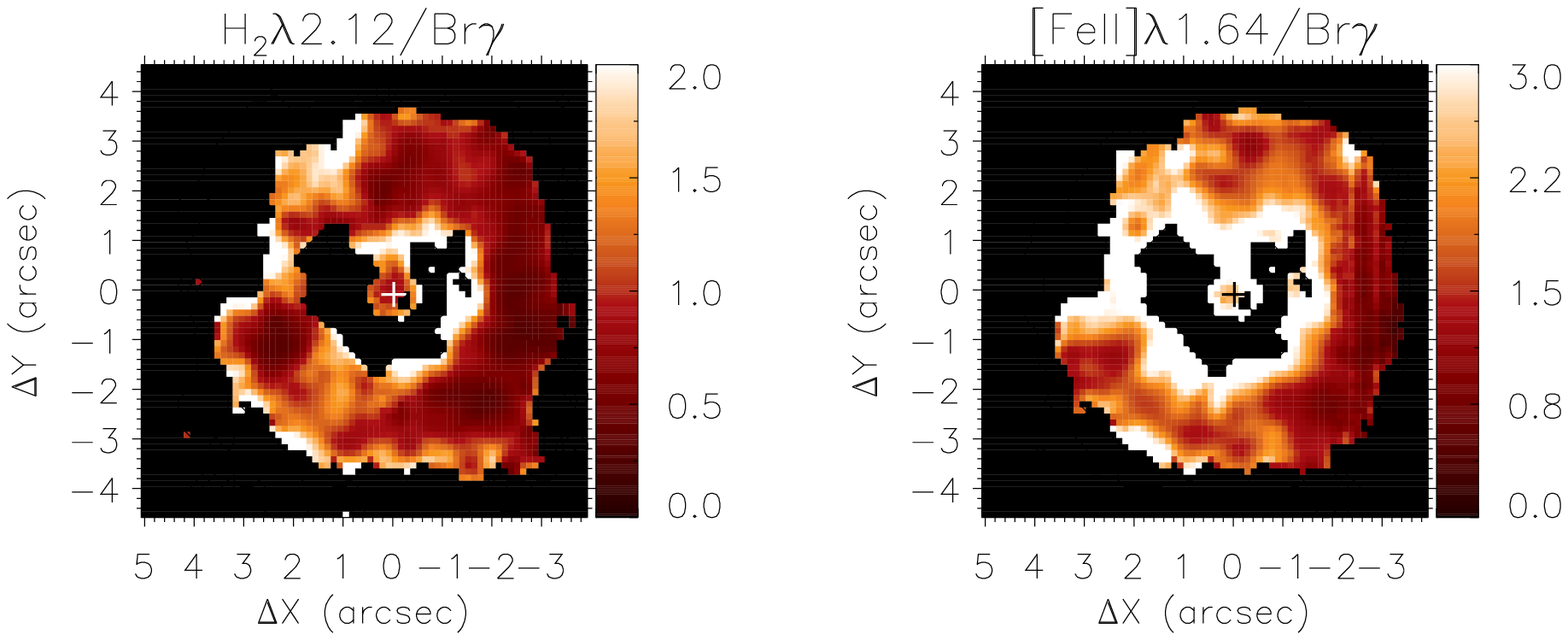} & 
\includegraphics[scale=0.48]{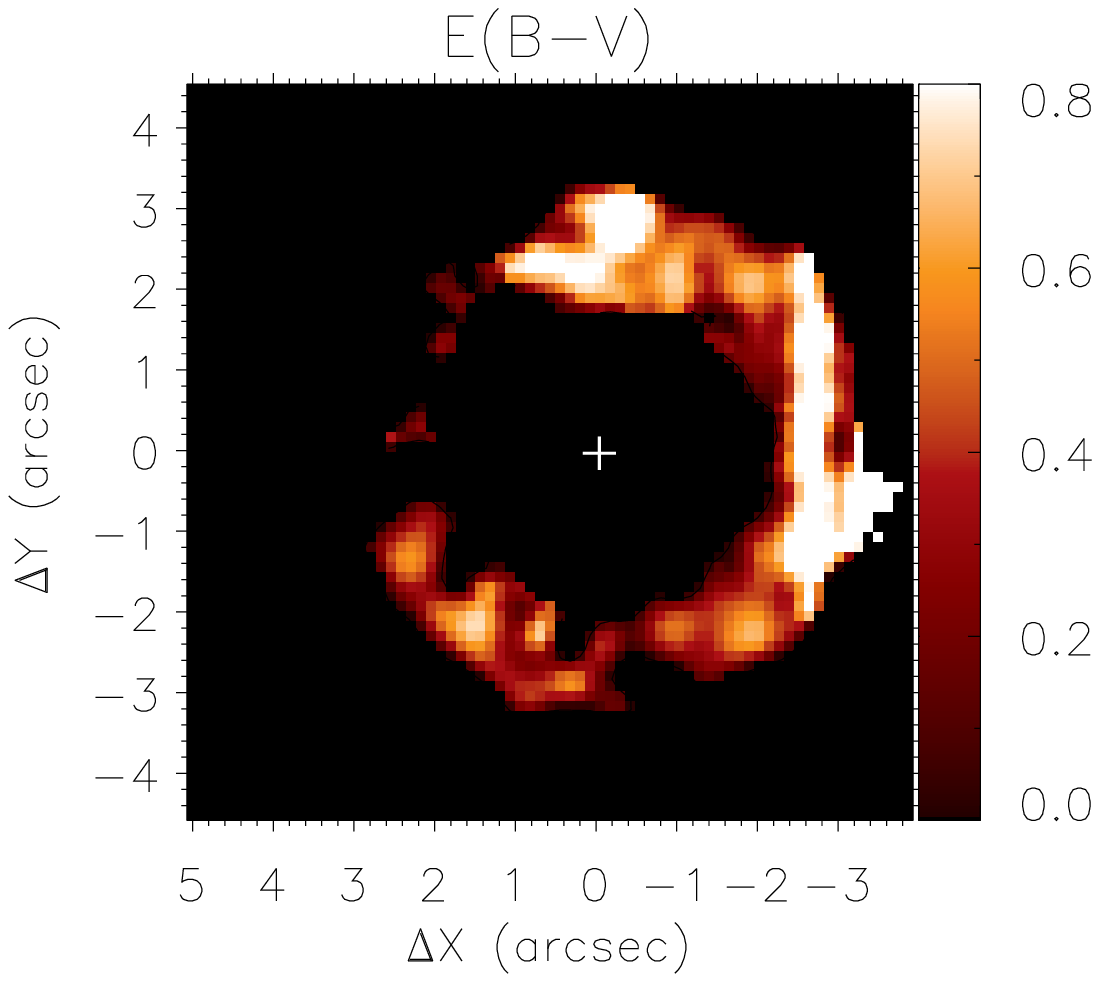}
\end{tabular}
\caption{\hml/\br\ and \feh/\br\  emission-line ratios and reddening map. The central cross marks the position of the nucleus and black locations are masked regions due to the non-detection of the emission lines or due to bad fits of their profiles. } 
\label{ratio}
\end{figure*}

In Figure~\ref{ratio} we present the \hml/\br\ and \feh/\br\ emission-line ratio maps, useful to investigate the excitation mechanisms of the \hm\ and \feii\ emission lines \citep[e.g.][]{colina15,rogerio13,dors12,ardila05,ardila04,reunanen02}. Both ratios present the smallest values at the locations of the star-forming clumps in the circumnuclear ring. Larger values are seen between the ring and the nucleus, but the uncertainties are high due to low \br\ emission. Typical values at the ring are \hml/\br$\sim$0.3 and \feh/\br$\sim$0.7, and, at the nucleus, $\sim1.4$ and $\sim3$, respectively. A reddening map obtained from Pa$\beta$/Br$\gamma$ is shown in the right panel of Fig.~\ref{ratio}, with values ranging from 0 to 0.8, but most locations having values smaller than 0.4. As the extinction in the near-IR is thus small, the fluxes and flux ratios were not corrected for reddening.

\begin{figure*}
\centering
\includegraphics[scale=0.9]{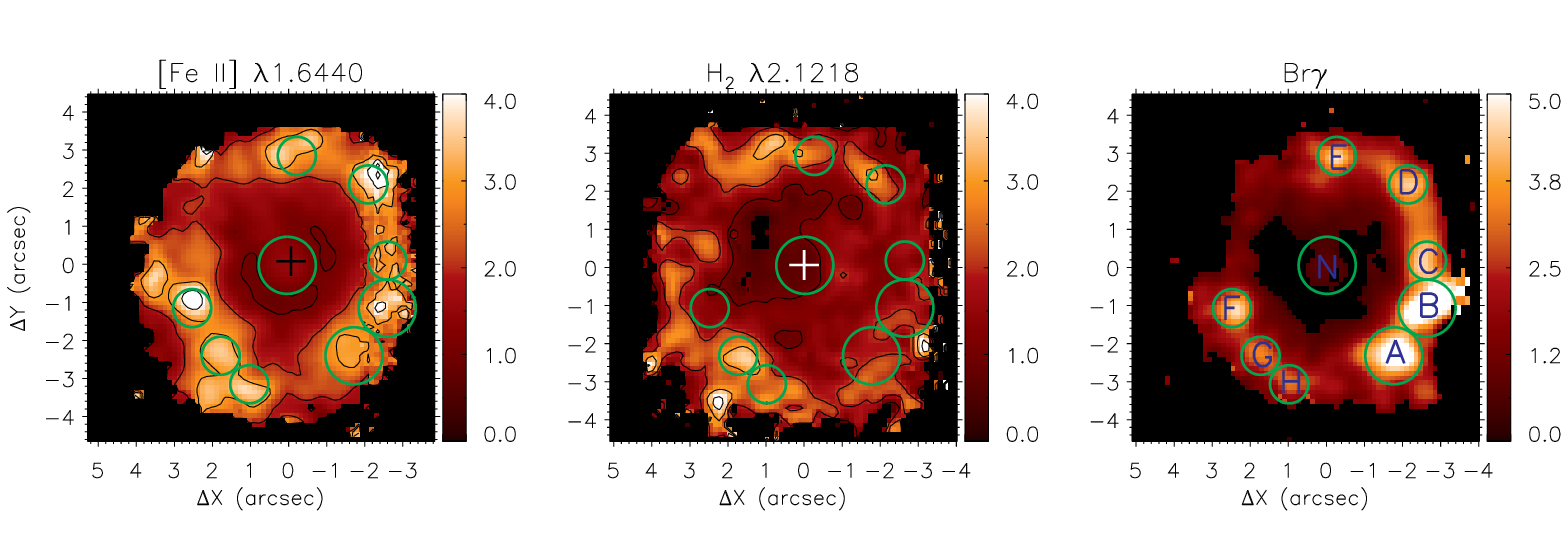}
\caption{Equivalent width maps in $\AA$ for the \feh, \hml\ and \br\ emission lines. The central cross marks the position of the nucleus and black locations are masked regions due to the non-detection of the emission lines or due to bad fits of their profiles. The circles in the \br\ map shows the regions used to extract the spectra of the CNSFRs and obtain the quantities listed in Table~\ref{tab-flux}. } 
\label{eqw}
\end{figure*}






Figure~\ref{eqw} shows maps for the EqWs of the \feii1.64, \hml\ and \br\ emission lines. The highest values of EqW at the ring reach 5\,$\AA$ for \br\ and 4\,$\AA$ for the \hm\ and \feii\ emission lines. 
Several knots attributed to CNSFRs are clearly observed in the \br\ EqW map which we used to extract the spectrum for each region. The EqW values we have measured are low when compared with those expected for young ($<$10 Myr) stellar populations, clearly indicating a large contribution of old stars to the near-IR continuum.

We have calculated the Equivalent Widths (EqWs) of the  \br, \pb, \feii\ and \hml\ emission lines for regions labeled from 
from A to F along the ring as well as for the nucleus and show their values, together with those of the emission-line fluxes in Table~\ref{tab-flux}. The spectra of these regions were extracted within apertures of 0\farcs75 for the nucleus and CNSFR A and 0\farcs50 for the remaining regions. The apertures were chosen to be larger than the seeing and to include most of the \br\ flux of each region.

\begin{table*}
\small
\caption{Emission line fluxes and Equivalent Widths for the CNSFRs in NGC\,4303. Fluxes are shown in units of ${\rm 10^{-16} erg s^{-1} cm^{-2}}$ and the Equivalent Widths in $\AA$. }
\vspace{0.3cm}
\begin{tabular}{l c c c c c c c c c c}
\hline

 Region & N & A& B & C & D& E& F& G& H \\
\hline
 R      &0\farcs75& 0\farcs75& 0\farcs75& 0\farcs50&  0\farcs50& 0\farcs50&  0\farcs50&  0\farcs50&   0\farcs50 \\
\hline
$F_{Br\gamma}$ & 5.2$\pm$ 3.7   & 7.1$\pm$ 0.8   & 7.9$\pm$ 1.0   & 3.3$\pm$ 0.5   & 2.6$\pm$ 0.5   & 2.4$\pm$ 0.5   & 2.8$\pm$ 0.5   & 1.9$\pm$ 0.3   & 1.6$\pm$ 0.4  \\
$EqW_{Br\gamma}$ & 0.7$\pm$ 0.5   & 4.4$\pm$ 0.6   & 4.7$\pm$ 0.7   & 4.0$\pm$ 0.6   & 3.9$\pm$ 0.8   & 3.5$\pm$ 0.7   & 3.7$\pm$ 0.7   & 2.9$\pm$ 0.5   & 2.7$\pm$ 0.7  \\
\hline
$F_{Pa\beta}$ & --        &44.2$\pm$ 8.9   &42.6$\pm$ 8.1   &18.0$\pm$ 3.8   &15.7$\pm$ 4.7   &12.9$\pm$ 4.7   &17.7$\pm$ 4.0   &14.6$\pm$ 4.1   &14.5$\pm$ 5.2  \\
$EqW_{Pa\beta}$ & --      &12.3$\pm$ 3.3   &11.9$\pm$ 3.3   & 9.5$\pm$ 2.4   & 9.8$\pm$ 3.3   & 8.2$\pm$ 3.1   & 9.2$\pm$ 2.4   & 9.8$\pm$ 3.5   &11.0$\pm$ 4.8  \\
\hline
$F_{[FeII]1.64}$  & 18.9$\pm$ 7.1   & 7.5$\pm$ 1.7   & 7.1$\pm$ 1.8   & 3.5$\pm$ 1.2   & 4.3$\pm$ 1.4   & 3.6$\pm$ 1.3   & 4.7$\pm$ 1.0   & 3.5$\pm$ 0.8   & 2.9$\pm$ 0.9  \\
$EqW_{[FeII]1.64}$&  1.4$\pm$ 0.5   & 3.0$\pm$ 0.7   & 3.1$\pm$ 0.9   & 3.0$\pm$ 1.1   & 3.8$\pm$ 1.4   & 3.2$\pm$ 1.3   & 3.7$\pm$ 1.0   & 3.3$\pm$ 0.8   & 3.1$\pm$ 1.1  \\
\hline
$F_{[FeII]1.25}$  & 34.3$\pm$ 8.3   &13.5$\pm$ 2.2   &12.6$\pm$ 2.4   & 6.5$\pm$ 1.4   & 7.2$\pm$ 1.3   & 5.3$\pm$ 1.4   & 7.7$\pm$ 1.2   & 6.3$\pm$ 0.9   & 5.6$\pm$ 1.2  \\
$EqW_{[FeII]1.25}$&   1.3$\pm$ 0.3   & 3.0$\pm$ 0.6   & 2.8$\pm$ 0.7   & 2.8$\pm$ 0.7   & 3.5$\pm$ 0.9   & 2.7$\pm$ 0.8   & 3.2$\pm$ 0.7   & 3.3$\pm$ 0.6   & 3.1$\pm$ 0.8  \\
\hline
$F_{H_2 2.12}$   &   7.5$\pm$ 3.6   & 3.5$\pm$ 0.8   & 3.4$\pm$ 0.7   & 1.5$\pm$ 0.5   & 1.9$\pm$ 0.4   & 1.8$\pm$ 0.3   & 1.3$\pm$ 0.4   & 2.2$\pm$ 0.3   & 1.8$\pm$ 0.4  \\
$EqW_{H_2 2.12}$&   0.9$\pm$ 0.4   & 2.1$\pm$ 0.5   & 1.8$\pm$ 0.4   & 1.8$\pm$ 0.6   & 2.7$\pm$ 0.6   & 2.4$\pm$ 0.5   & 1.6$\pm$ 0.5   & 3.1$\pm$ 0.5   & 2.8$\pm$ 0.6  \\ 
\hline
\end{tabular}
\label{tab-flux}
\end{table*}

\subsection{Stellar and gas kinematics}

\begin{figure*}
\centering
\includegraphics[scale=0.9]{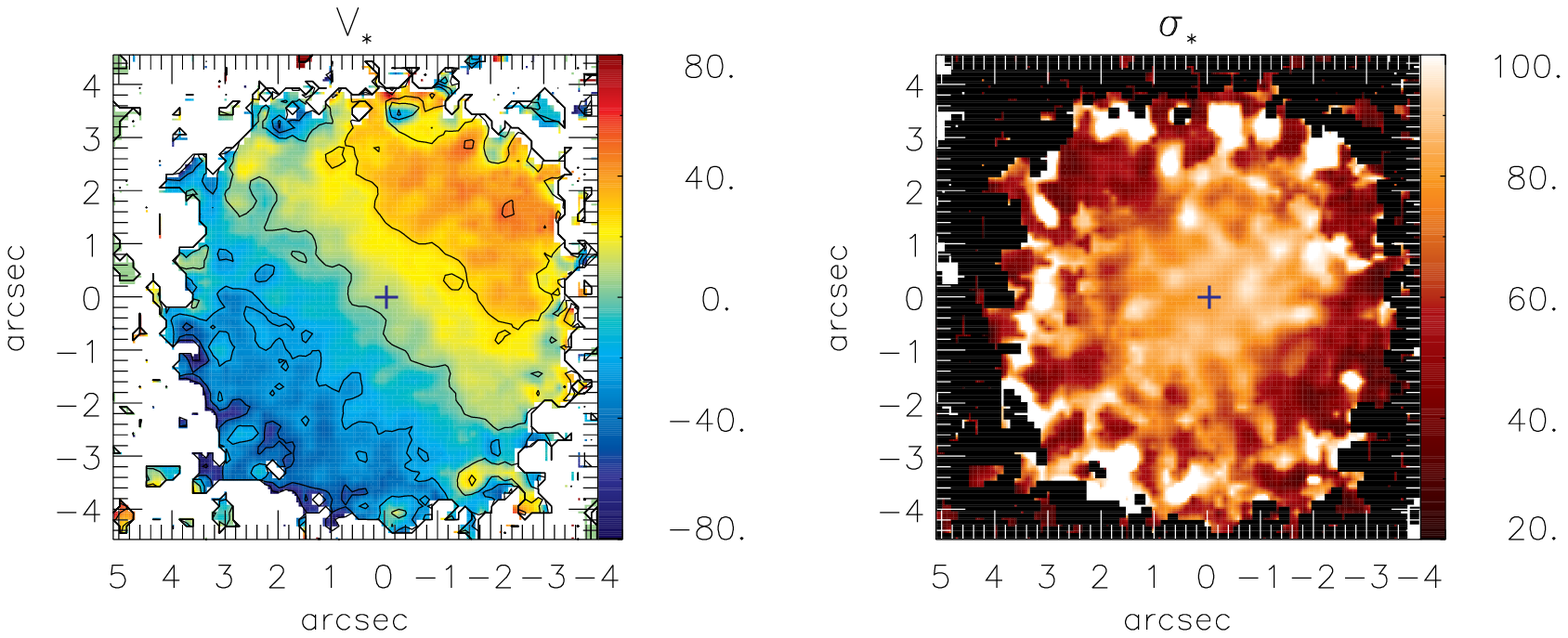}
\caption{Stellar velocity ($V_*$) field (left)  and velocity dispersion ($\sigma_*$) map (right)  obtained from the fit of the CO absorption band heads at 2.3\,$\mu$m using the pPXF routine. The central cross marks the position of the nucleus and white/black locations in $V_*/\sigma_*$ maps are masked regions due to bad fits of their profiles. The color coded maps are given in units of km\,s$^{-1}$.} 
\label{stel}
\end{figure*}

We used the 
penalized pixel-fitting (pPXF) method of \citet{cappellari04} to fit the the CO absorption band heads at 2.3\,$\mu$m and obtain the stellar velocity ($V_*$) field, velocity dispersion ($\sigma_*$) and higher order Gauss-Hermite moments ($h_3$ and $h_4$) as well as  their uncertainties.  
We used the library of 60 stellar spectra  of early-type stars from \citet{winge09} as template spectra and fitted the galaxy spectra in the range from 2.29 to 2.41\,$\mu$m. The spectral resolution of the stellar spectra is better than the resolution of the galaxy spectra, thus we degraded the template spectra to the same resolution of the SINFONI observations by convolving the spectra with a Gaussian function to get reliable $\sigma_*$ measurements. 

In Figure~\ref{stel} we present the stellar velocity field in the left panel and the $\sigma_*$ map in the right panel. Black regions in these maps correspond to locations 
where the uncertainties in $V_*$ and/or $\sigma_*$ are larger than 30~\kms.  The systemic velocity of the galaxy, derived in Sec.~\ref{disc-stel}, was subtracted from the measured velocities and the color bars are shown in units of \kms.  The stellar velocity field is consistent with rotation in a disc with the major axis oriented along the position angle PA$\sim135/315^\circ$ and with a projected (along the line of sight) velocity amplitude of about 70~\kms. The $\sigma_*$ map shows values ranging from $\sim$20 to $\sim$100~\kms, with the lowest values observed in a ring with radius of $\sim$2.5--3$^{\prime\prime}$ surrounding the nucleus, corresponding to the location of the CNSFRs. 
The largest $\sigma_*$ values are observed in patchy regions that appear just beyond or surrounding the star-forming clumps, that we attribute to the velocity dispersions of the stellar bulge.



\begin{figure*}
\centering
\includegraphics[scale=1]{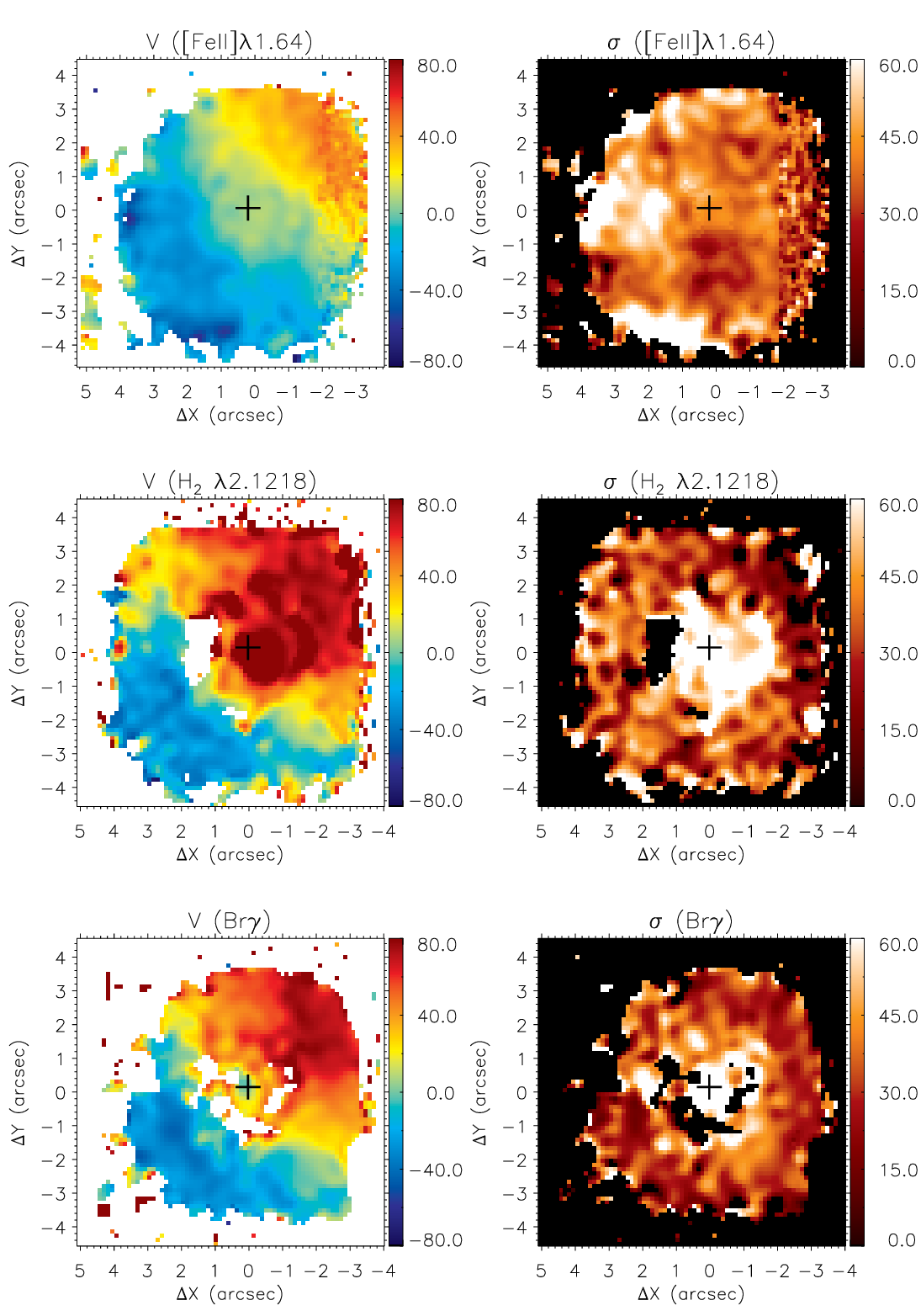}
\caption{Velocity ($V$) fields (left) and velocity dispersion ($\sigma$) maps (right) for the \feh\ (top), \hml\ (middle) and \br\ emission lines. The central cross marks the position of the nucleus and white/black locations in $V/\sigma$ maps are masked regions due to the non-detection of the emission lines or due to bad fits of their profiles.  The color bars are shown in units of km\,s$^{-1}$.} 
\label{gas-kin}
\end{figure*}

Figure~\ref{gas-kin} presents the velocity fields and velocity dispersion ($\sigma$) maps for the \feii1.64, \br\ and \hml\ emission lines. We do not show the maps for the \feii1.25 and \pb\ emission lines because they are very similar to those for the  \feii1.64 and \br\ lines, respectively, and have lower signal-to-noise ratios. 
All  velocity fields present a similar rotation pattern as seen for the stars with the orientation of the line of nodes ($\sim$130$^\circ$) and the projected velocity amplitude (70\,\kms) in good agreement with the values previously derived from the H$\beta$ emission line \citep{colina97}.
The $\sigma$ maps
show small values over the whole field of view with the highest values reaching only 60~\kms.  The \feii\ map shows the highest values to the east and south-south-east of the nucleus and the \hm\ and \br\ present the highest values at the nucleus. The uncertainties in $\sigma$ and velocity are smaller than 15~\kms\ for all emission lines at most locations. For the nucleus, the uncertainties for the \hm\ and \br\ are larger due to the low $snr$ for these lines, with typical uncertainties of 25~\kms.  Uncertainties of about 20~\kms\ are observed for the \feii\ at locations between 2-3$^{\prime\prime}$ west (right side of the field of view)  due to a lower exposure time at these locations (effect of the spatial dithering).

\section{Discussion}

\subsection{(Circum)nuclear stellar and gas distribution: from spirals to ring structure?}\label{distributions}

\begin{figure}
\includegraphics[scale=0.8]{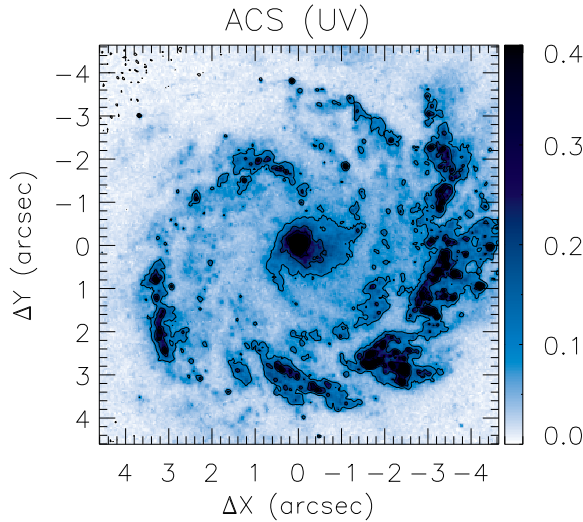} 
\caption{UV HST image of NGC\,4303 previously discussed by \citet{colina02}.} 
\label{hst}
\end{figure}

UV and near-IR Hubble Space Telescope (HST) images of NGC\,4303 show a nuclear spiral with several knots of star forming clusters, as seen in Fig.\,\ref{hst} and a weak nuclear stellar bar \citep{colina00}, respectively.  The nuclear spiral is not evident in the near-IR emission line flux distributions shown in Fig.~\ref{fluxmaps}, where the structure observed can be better described as a ring of CNSFRs with radius of about 2.5--3.2$^{\prime\prime}$ (200--250\,pc). The CNSFRs seem to correspond to complexes of the strongest  star-forming clusters observed in the UV HST image. This ring is coincident with the inner-Inner Lindblad resonance, as derived by \citet{schinnerer02} from the observations of the CO cold molecular gas, explaining why the gas has accumulated in the ring, giving origin to the star clusters.


The emission-line flux distributions presented in Fig.~\ref{fluxmaps} show distinct structures for the different emission lines.  The strongest emission for the \feii\ and \hm\ lines is from the nucleus, while the hydrogen recombination lines show emission predominately from the CNSFRs.
The \feii\ shows an elongation to $\approx$\,1.5" (120\,pc) north-east, where an outflow has been observed in [O\,{\sc iii}]  line emission  \citep{colina99}. No emission associated to this outflow is observed in the \hm\ map, probably due to the dissociation of the \hm\ molecule by the AGN emission, as commonly seen for Seyfert galaxies \citep[e.g.][]{n1068-exc,sb09}.  The nuclear \hm\  emission shows instead an elongation towards the west, the same orientation observed in the cold molecular gas emission, connecting the nucleus with the circumnuclear ring \citep{schinnerer02}.

In the ring, although the flux distribution in \feii\ is similar to those of the hydrogen recombination lines, the brightest \pb\ and \br\ regions (A and B) are located south-west of the nucleus, while for \feii\ the brightest region is an unresolved knot (F) to the south-east.
This difference is consistent with the scenario in which regions A and B are younger than region F, that has stronger \feii\ emission due excitation by shocks from supernovae (SN), contributing more at a later age (see discussion below).

The hot \hm\ emission is strongest in the east part of the ring, at locations that seem to be anti-correlated with the locations of strongest \pb, \br\ and \feii\ emission, that coincide also with the UV knots of recent star formation. In previous studies by our group \citep[e.g.][]{sb09} we concluded that the temperature of the region emitting hot H$_2$ is $\approx$\,2000K, and that the excitation mechanism was thermal, as a result of heating by either X-rays from AGN or shocks from both AGN or SNe. In the ring, the eastern H$_2$ knots of  emission probably originate in regions that are even older than the eastern knot that emits \feii\ and \br\ lines. These regions would still be hot enough to produce H$_2$ but not enough to produce \br\ and \feii\ emission (that require 10,000--15,000\,K). 

Finally, the K-band image in Fig.~\ref{fluxmaps} show a small elongation to the north-east and south-west, visible in the second isocontour which is similar to that observed in the higher resolution HST image \citep{colina00} and attributed to a weak nuclear stellar bar.

\subsubsection{The origin of the nuclear emission}\label{disc-nuc-emission}

The origin of the nuclear gas emission in NGC\,4303 is still an open question, but previous works have suggested that the ionizing source is a combination of a low-luminosity AGN and a young massive stellar cluster  \citep{colina97,colina02,colina99}. 

The near-IR emission-line ratios can be used to further investigate this point. The line-ratio values from the nucleus can also be compared with those from the CNSFRs. For the nucleus, we obtain  \hm/\br$=1.4\pm1.2$  and \feii$\lambda1.64$/\br\ =$3.6\pm2.8$ using the values listed in Table~\ref{tab-flux}. These values are typical of Seyfert nuclei, but also of excitation by shocks from SNe  \citep{reunanen02,ardila04,ardila05,colina15}. They are also significantly higher than those observed in the ring, for example, from  regions A and B (\hm/\br$\sim$0.4  and \feii$\lambda1.64$/\br\ $\sim$1.0) where the presence of young (2.5--7.5\,Myr) clusters dominating the ionization is well established from UV spectroscopy \citep{colina02}.
Thus, the near-IR diagnostics also suggest the presence of an AGN at the nucleus, confirming previous results that the nuclear emission has a composite nature. Alternatively, a SNe-dominated 8--40~Myr star forming cluster could produce such enhanced line ratios, but there is no evidence so far of a cluster  with this age at the nucleus, only of a younger one  ($\approx$4 Myr old). Moreover, the fact that the \feii\ emission at the nucleus is extended along the ionized [OIII] outflow, suggests that the nuclear emission of \feii\ and the \hm\ could be mainly due to X-rays emitted by the AGN \citep{jimenez03} and/or shocks in the outflow. 

\subsubsection{The circum-nuclear star forming regions}\label{cnsfr-disc}
 
\citet{colina97} reported the presence of a circumnuclear spiral structure with several UV-bright knots identified as young (few Myr old) star-forming regions \citep{colina99}. The CNSFRs observed in the flux maps of Fig.~\ref{fluxmaps} and EqW maps of Fig.~\ref{eqw} show a structure similar to that seen in the UV continuum and optical emission-line flux maps \citep{colina99}, but with a ring rather than a spiral structure.  (We note that the region identified as A in Fig.~\ref{eqw} corresponds to regions G and F of \citet{colina97}). As discussed in Sec.~\ref{distributions}, the location of the ring agrees with that of the inner Lindblad resonance and therefore it may be a stable structure.

Assuming that the CNSFR ring is circular and located in the plane of the galaxy, we can use the observed geometry to derive the inclination of the ring relative to the plane of the sky. In the \br\ EqW map (Fig.~\ref{eqw}) the CNSFR shows an elliptical shape with major axis of $a\sim6^{\prime\prime}$ oriented along PA$\approx$135$^\circ$ and a minor axis of $b\sim5^{\prime\prime}$. The resulting inclination of the ring is $i={\rm acos}(b/a)\approx33^\circ$, which is similar to the inclination of large scale disk of NGC\,4303 ($i\approx27^\circ$). 

The \hm/\br\ and \feii/\br\ maps show small values in the CNSFR (see Fig.~\ref{ratio}). \feii/\br\ is smaller than 2 for all star-forming regions, being consistent with values expected for young-stellar clusters \citep{colina15}. The \hm/\br\ ratio for most star-forming regions is smaller than 0.6, as expected for a young cluster, with exception of the regions G and H (identified in Fig.~\ref{eqw}) with higher values (of $\sim 1$), suggesting an additional mechanism for the excitation of the \hm\ emission there, such as shocks from  supernovae explosions or diffuse X-ray emission heating the gas and exciting the \hm\ molecule.

The integrated emission-line fluxes given in Table~\ref{tab-flux} have been used to derive a number of physical properties of the star-forming clumps in the CNSFR that are listed in Table~\ref{tab-pars}, as follows. The values of the rate of ionizing photons $Q[H^+]$ and Star Formation Rate ($SFR$) were derived under the assumptions of continuous star formation, and should be considered only a proxy of these parameters. 
 
The emission rate of ionizing photons for each star-forming region was obtained following \citet{n7582}, using: 
\begin{equation}
\left(\frac{Q[H^+]}{\rm s^{-1}}\right)=7.47\times10^{13}\,\left(\frac{L_{Br\gamma}}{\rm erg\,s^{-1}}\right)
\label{q}
\end{equation}
and the star formation rate ($SFR$)  using \citep{kennicutt98}:
\begin{equation}
\left(\frac{SFR}{\rm M_\odot yr^{-1}}\right)=8.2\times10^{-40}\,\left(\frac{L_{Br\gamma}}{\rm erg\,s^{-1}}\right),
\end{equation}
where $L_{Br\gamma}$ is the \br\ luminosity, under the assumption of continuous rate. 

In order to estimate the mass of stars of each CNSFR, we need to assume an age for the region. Detailed Spectral Energy Distribution (SED) fitting of the HST UV spectra of the nucleus and of the two brightest UV circumnuclear regions  \citep{colina02} led to the conclusion that they are best reproduced by instantaneous star-forming bursts of age about 4 Myr. This age was thus used to estimate the mass of each region under the assumption of instantaneous star formation using the {\sc starburst 99} code \citep{leitherer99,leitherer10,leitherer14,vazquez05} and assuming the same Initial Mass Function (IMF) parameters as in \citet{colina02} -- i.e. a Salpeter IMF, with mass of the stars between 1 and 100\,M$_\odot$ and solar metallicity. Considering these parameters, a 4 Myr old cluster with mass $1\times10^5\,$M$_\odot$ will produce a rate of ionizing photons ($<$13.6 eV) of  log$Q_M$ $\sim$ 51.1 photons/s. Therefore, the mass of the young star-forming regions can be obtained from the observed log$Q$ listed in  Table~\ref{tab-pars} as $M=1\times10^5\,{\rm M_\odot}\,Q/Q_M$. The resulting mass for the clusters are shown in Table~\ref{tab-pars} and are in the range 0.30--1.45\,$\times10^5$\,M$_\odot$.  
For more evolved regions (e.g., with ages $\sim$ 10 Myr old instead of 4 Myr), the masses will increase by a factor of one hundred as the ionizing radiation is a steep function of age for an instantaneous burst.

The mass of ionized ($M_{\rm H\,II}$) gas can be derived as  \citep[e.g.][]{osterbrock06,sb09}:

\begin{equation}\label{mhii}
 \left(\frac{M_{\rm H\,II}}{M_\odot}\right)=3\times10^{19}\left(\frac{F_{\rm Br\gamma}}{\rm erg\,cm^{-2}\,s^{-1}}\right)\left(\frac{D}{\rm Mpc}\right)^2\left(\frac{N_e}{\rm cm^{-3}}\right)^{-1},
\end{equation}
where $D$ is the distance to the galaxy, $F_{\rm Br\gamma}$ is the Br$\gamma$ flux and $N_e$ is the electron density. We have assumed an electron temperature of 10$^4$K and density of $N_e=500$\,cm$^{-3}$. The mass of hot molecular gas is estimated using \citep[e.g.][]{scoville82,n4051,n1068-exc}:
 
\begin{equation}\label{mhm}
 \left(\frac{M_{H_2}}{M_\odot}\right)=5.0776\times10^{13}\left(\frac{F_{H_{2}\lambda2.1218}}{\rm erg\,s^{-1}\,cm^{-2}}\right)\left(\frac{D}{\rm Mpc}\right)^2,
\label{mh2}
\end{equation}
where $F_{H_{2}\lambda2.1218}$ is the H$_2$ (2.1218$\mu$m) emission-line flux and the assumptions of local thermal equilibrium and excitation temperature of 2000\,K were used.


The resulting values of the ionized and hot molecular gas masses for the nucleus and for each CNSFR are also listed in Table~\ref{tab-pars}. In summary, assuming a typical age of about 4 Myr for the star-forming clumps,
the stellar masses in young stars is in the range 3--10\,$\times$\,10$^4$\,M$_{\odot}$, the ionized gas mass is 
about 10 times lower and the mass of hot molecular gas is in the range 3--4\,$\times$\,10$^{-5}$\,M$_{\odot}$.
The ratio between the masses of ionized and hot molecular gas are between 1000 and 2800, with the lower ratios observed for the regions G and H. For the nucleus, $M_{HII}/M_{H2}\approx830$. These ratios are similar to those in the central region of Seyfert galaxies \citep[e.g.][]{n1068-exc,sb09}, although distinct scenarios are expected for the heating of the gas. In AGNs, the H$_2$ emission is usually due to thermal processes, i.e. heating of the gas by shocks or X-rays from the AGN, while for the CNSFRs, young stars play an important role in the excitation of the H$_2$. Finally, the $Q[H^+]$ and equivalent $SFR$ derived for the CNSFR of NGC\,4303 are in good agreement with those obtained for other CNSFRs in nearby  galaxies characterized by a moderate star-forming regime \citep[e.g.][]{falcon-barroso14,dors08,shi06,galliano08,wold06}.


\begin{table*}

\caption{Physical parameters of the CNSFRs in NGC\,4303. The location of each region is indicated in Fig.~\ref{eqw}. $Q$[H$^+$]: ionizing photons rate, $SFR$: equivalent star formation rate under the assumption of continuous star formation following \citet{kennicutt98}, $M$: mass of the cluster assuming a cluster instantaneously formed with age 4~Myr, $M_{HII}$: mass of ionized gas, $M_{H2}$: mass of hot molecular gas, $\sigma_l$: velocity dispersion for the line $l$.}
\vspace{0.3cm}
\begin{tabular}{l c c c c c c c c c c}
\hline

 Region & N& A& B & C & D& E& F& G& H \\
\hline
log $Q$[H$^+$]\,(s$^{-1}$) & 51.1$\pm$ 0.12   &51.2$\pm$ 0.02   &51.3$\pm$ 0.02   &50.9$\pm$ 0.02   &50.8$\pm$ 0.03   &50.7$\pm$ 0.03   &50.8$\pm$ 0.03   &50.6$\pm$ 0.03   &50.6$\pm$ 0.04  \\

$SFR\, ({10^{-2} \rm M_ \odot yr^{-1}})$ &  1.3$\pm$ 0.31   & 1.8$\pm$ 0.07   & 2.0$\pm$ 0.09   & 0.8$\pm$ 0.04   & 0.7$\pm$ 0.04   & 0.6$\pm$ 0.04   & 0.7$\pm$ 0.04   & 0.5$\pm$ 0.03   & 0.4$\pm$ 0.03  \\

$M ({\rm 10^5 M_ \odot})$&  1.0$\pm$0.30 & 1.3$\pm$0.06 & 1.5$\pm$0.06 & 0.6$\pm$0.03 & 0.5$\pm$0.03 & 0.4$\pm$0.03 & 0.5$\pm$0.04 & 0.4$\pm$0.02 & 0.3$\pm$0.02 \\ 

$M_{HII}\, (10^3{\rm M_ \odot})$&  8.2$\pm$ 1.9   &11.1$\pm$ 0.4   &12.4$\pm$ 0.5   & 5.1$\pm$ 0.3   & 4.0$\pm$ 0.3   & 3.7$\pm$ 0.2   & 4.4$\pm$ 0.3   & 2.9$\pm$ 0.2   & 2.5$\pm$ 0.2  \\

$M_{H2}\, ({\rm M_ \odot})$&  9.8$\pm$ 1.6   & 4.7$\pm$ 0.3   & 4.4$\pm$ 0.3   & 2.0$\pm$ 0.2   & 2.5$\pm$ 0.2   & 2.4$\pm$ 0.1   & 1.7$\pm$ 0.2   & 2.9$\pm$ 0.1   & 2.4$\pm$ 0.2  \\
 
$\sigma_{FeII}\, ({\rm km\,s^{-1}}$) & 44.6$\pm$10.4   &40.1$\pm$ 6.1   &37.9$\pm$ 6.7   &44.3$\pm$ 9.2   &45.5$\pm$ 9.2   &49.5$\pm$10.3   &52.9$\pm$ 6.5   &43.2$\pm$ 6.1   &58.9$\pm$ 9.6  \\

$\sigma_{H2}\, ({\rm km\,s^{-1}}$) & 80.2$\pm$16.2   &49.6$\pm$ 5.5   &39.1$\pm$ 4.7   &40.0$\pm$ 7.3   &36.7$\pm$ 4.4   &36.7$\pm$ 3.6   &43.8$\pm$ 7.3   &43.5$\pm$ 3.1   &39.3$\pm$ 4.6  \\

$\sigma_{Br\gamma}\, ({\rm km\,s^{-1}}$) & 81.8$\pm$23.7   &40.4$\pm$ 2.6   &32.3$\pm$ 2.5   &33.7$\pm$ 3.0   &32.8$\pm$ 3.9   &33.7$\pm$ 3.9   &30.9$\pm$ 3.5   &34.2$\pm$ 3.6   &29.3$\pm$ 4.6  \\
\hline
\end{tabular}
\label{tab-pars}
\end{table*}

\subsubsection{The \feii/\br\ vs. \hm/\br\ diagnostic diagram}

\begin{figure}
\centering
\includegraphics[scale=0.5]{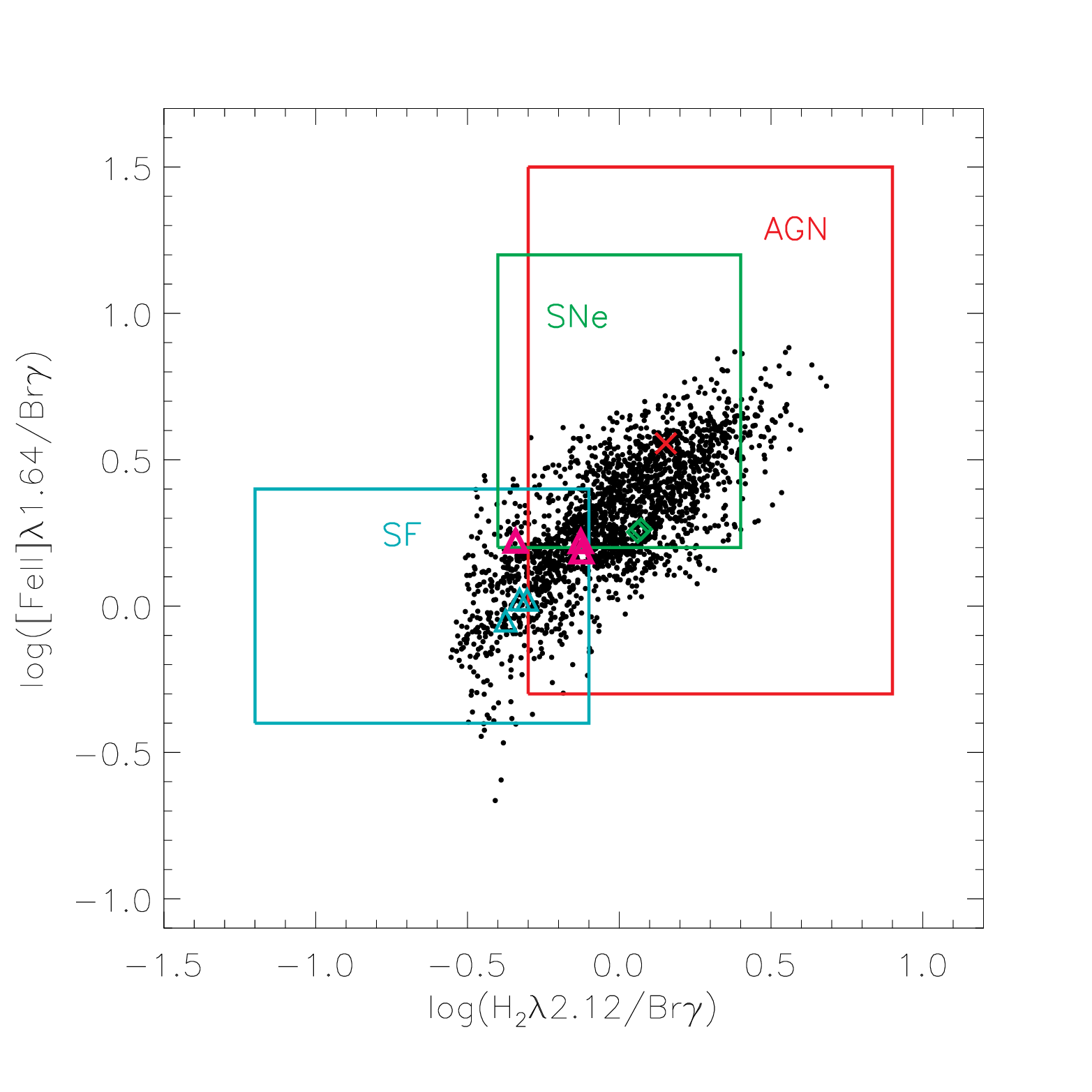}
\caption{\feh/\br\ vs. \hml/\br\ diagnostic diagram for the central region of NGC\,4303. The black points corresponds to all the spaxels for which we were able to measure both line ratios. The red cross represents the nucleus, the green diamonds are for regions G and H and the pink(blue) triangles represent the ratios for regions D to F (A to C), as labeled  in Fig.~\ref{eqw}. The limits corresponding to excitation by young stars (SF), supernova remnants (SNe) and  AGN are from \citet{colina15}.} 
\label{diagn}
\end{figure}

In order to better study the gas excitation, we constructed the  \feii/\br\ vs. \hm/\br\ diagnostic diagram shown in Fig.~\ref{diagn}. This diagram shows a big overlap of regions corresponding to SNe and AGN, as some excitation processes, like shocks and thermal heating are common to both AGN and SN.
The black dots correspond to all observed spaxels and are mostly located in the region of SNe and/or AGN, with some points covering part of the star-formation (SF) region. This behavior is typical of luminous star-forming infrared galaxies and interpreted as due to excitation by combined effects of the AGN and stellar ionizing radiation plus shocks in AGN outflows, stellar winds and supernovae explosions \citep{colina15}.  

For the nucleus, the line ratios are high and typical of AGNs, while for the brightest regions in the UV (A to C) in the CNSFR, that are the  youngest (see discussion below) the line ratios are typical of star forming regions. The older star-forming regions in the CNSFR (D to H) are displaced towards the SNe region, with regions G and H presenting the most extreme ratios, likely associated to SNe explosions. The line ratios are thus compatible with an age difference between the east (older) and west (younger) side of the ring (see Sec. 4.1.4 for further discussion).


\subsubsection{How does the star formation proceed in the nucleus and CNSFRs ring?}

The star formation in the nuclear regions of early type spirals frequently occurs in ring-like structures with radius of 1 kpc or less. Examples include galaxies with luminous AGNs in their centers as well as many others with a weak AGN or star-forming nucleus \citep{genzel95,sb96a,sb96b,boker08,laan13a,laan13b,falcon-barroso14,laan15}. It is well established that bars in galaxies can drive large quantities of gas to the inner regions, accumulating into inner resonances, generating and maintaining subsequent star formation there \citep{combes85,heller96,buta96}.  Two scenarios have been considered so far to 
explain the star formation in circumnuclear rings:  the ``pop-corn" \citep{elmegreen94} and the ``pearls on a string" \citep{boker08}. As discussed in the Introduction, in the ``pop-corn" scenario the stellar clusters form at random positions with no age sequence, while in the ``pearls on a string" scenario, the clusters  are formed where the gas enter the rings and then age as they orbit the ring forming a string of aging clusters. 

Some previous studies favor the ``pearls-on-a-string"  scenario \citep{boker08,falcon-barroso14}, in galaxies with evidence of gas inflow into the ring \citep{laan13a}. However, there are also rings that appear to be a combination of the two scenarios \citep{laan15}. An  H$\alpha$ imaging survey of 22 galaxies with circumnuclear star-forming rings \citep{mazzuca08} has shown that about half of the rings show azimuthal age gradients as expected in the ``pearls-on-a-string" scenario, while the other half show no age pattern, have a flat age  distribution, or even a radial gradient.  Thus, the star formation in the (circum)nuclear regions of galaxies is far from clear. 

\begin{figure}
\centering
\includegraphics[scale=0.45]{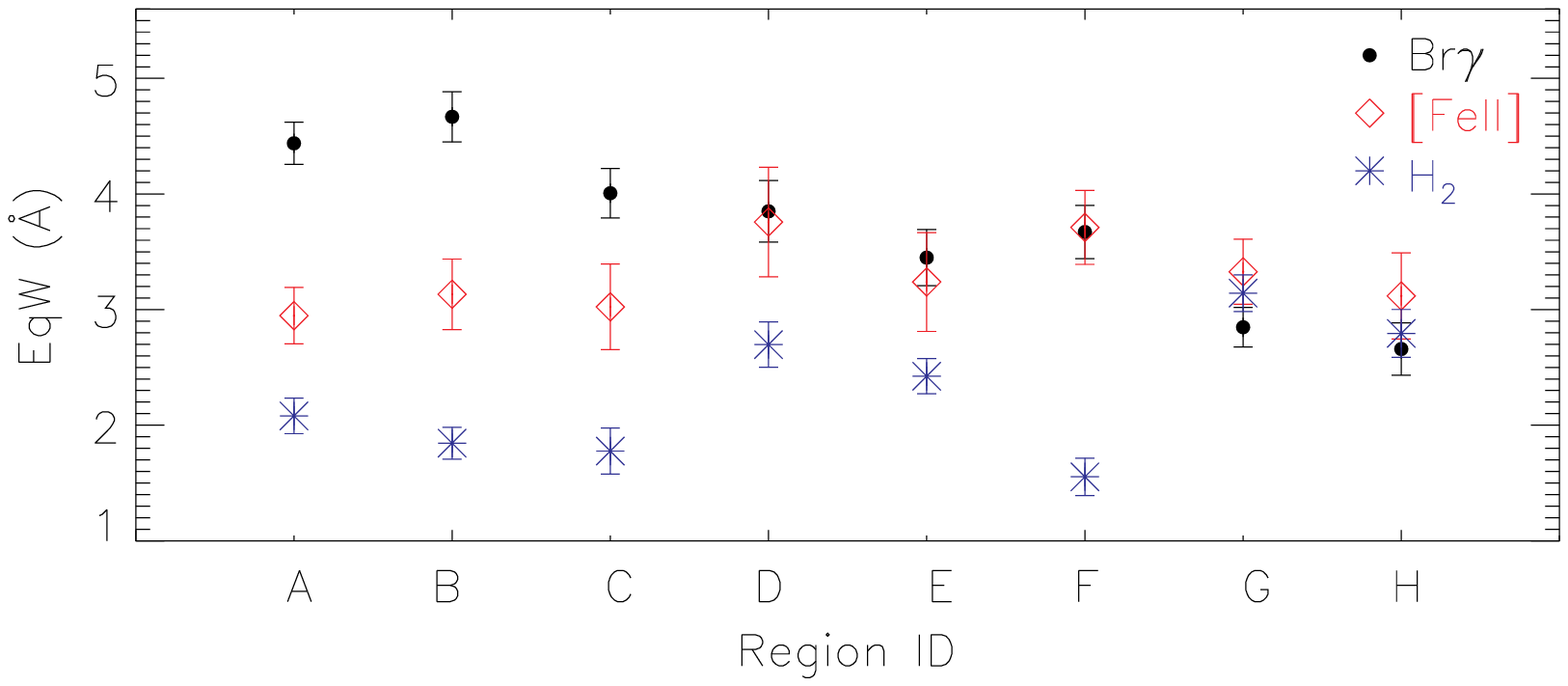}
\includegraphics[scale=0.45]{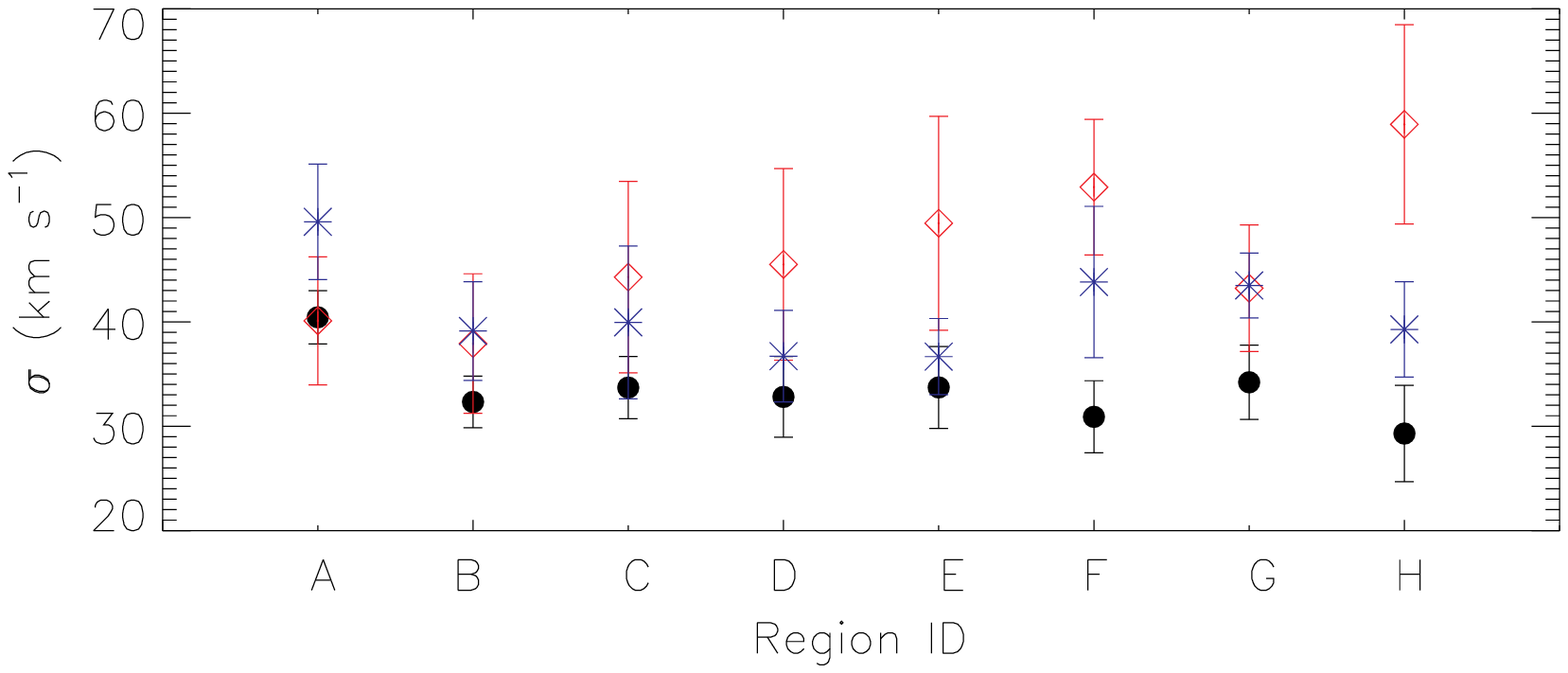}
\includegraphics[scale=0.45]{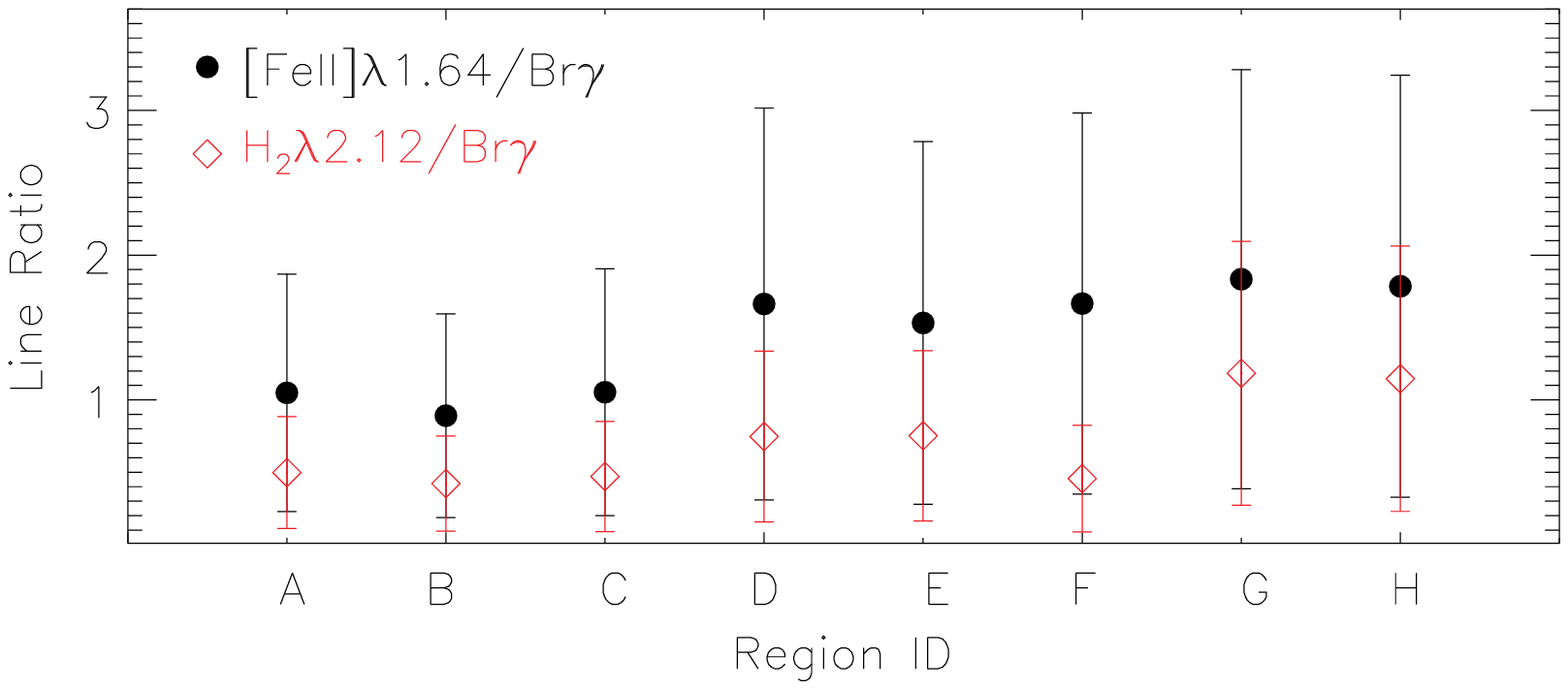}
\caption{Plots for the $EqW$ (top), $\sigma$ (middle) and line ratios (bottom) vs. region. } 
\label{cnsfr-plots}
\end{figure}

In order to investigate the best scenario for the  circum-nuclear star formation in NGC\,4303, we show in Fig.~\ref{cnsfr-plots} the sequence of values of the EqW and $\sigma$ for the \hm, \br\ and \feii\ emission lines, as well as the corresponding \feii/\br\ and \hm/\br for each CNSFR, labeled from A to H.
The top panel shows a small decrease in EqW values for \br, consistent with a small decrease in age from A to H. For the velocity dispersion (middle panel), no clear trend is observed for \br\ and \hm, while for the \feii, the $\sigma$ values show a small increase from regions A to H. This variation is also consistent with a small increase in age from regions A to H, as the presence of SNs in older regions can increase the \feii\ velocity dispersion. 
The bottom panel of Fig.~\ref{cnsfr-plots} show small positive gradients for both \hm/\br\ and \feii/\br\ line ratios between regions A and H, and can also be understood as due to an increase in age as shocks from supernovae contribute as an additional excitation source for the \hm\ and \feii\ as the regions age, while the evolution of massive stars away from the main sequence diminishes the amount of ionizing radiation, i.e. reducing the flux of \br.  This relative age dating is in agreement with previous conclusions \citep{colina00,colina02} that identified the east UV knots as older ($\sim$10--25 Myr) than the west knots ($\sim$2.5--7.5 Myr), therefore suggesting an age offset in the star-forming ring. Thus, relative age differences for the star-forming clumps are confirmed from independent tracers in the UV-optical and now the near-IR. 

However, although our measurements show a trend suggesting an age sequence along the ring (increasing from A to H), the trend is small and present some ups and downs, being equally consistent just with regions to the east being older than regions to the west. For example, in the case of \hm/\br\ instead of a smooth gradient, the observed variation is more consistent with regions G and H being older than the regions A--F. We conclude that the data do not confirm the ``pearls on a string" scenario neither the ``pop-corn'' scenario, as there is also no evidence for a random distribution of ages. We can only be sure of an age asymmetry, with the youngest regions to the west and oldest regions to the east.


One interesting new result that can be observed in Figure \ref{fluxmaps} is that the regions of strongest H$_2$ emission in the ring are approximately anti-correlated with the regions of strongest \br\ and \feii\ emission. While the latter coincide with the UV knots of recent star formation, the H$_2$ knots are observed mostly in the east part of the ring, but not coincident with the bright \br\ and \feii\ knot there. In previous results from our group \citep[e.g.][]{sb09} we concluded that the temperature of the region where hot H$_2$ is observed is $\approx$\,2000K, and that the mechanism producing the hot H$_2$ emission was thermal excitation as a result of heating either by X-rays from the AGN or shocks from SN. In the ring the dominant mechanism should be shocks from SN, thus the eastern knots of H$_2$ emission would be hot (2000\,K) regions associated with fading SN, being older than the eastern knot of \feii\ and \br\ emission. These regions would be hot enough to produce H$_2$ but not to produce \br\ and \feii\ emission (requiring 10,000--15,000\,K). 

Our results can be compared with those for the cold molecular gas \citep{schinnerer02}. The asymmetry in the star formation along the ring appears to follow the distribution of cold molecular gas, dominated in mass and surface density by the western spiral and gas lane \citep{schinnerer02}. At the resolution of the CO observations (2$^{\prime\prime}$), the large concentration of molecular gas in the spiral seems to be located just outside the CNSFR, suggesting therefore not a direct correspondence between the regions were the star formation is taking place now, and the regions with the highest molecular gas densities (Figure 10 in \citet{schinnerer02}). In addition, the east side of the ring appears to be forming very few stars now (as judged from the \br\ and UV emission), and is almost devoid of cold molecular gas, in particular the northeast region \citep{schinnerer02}. On  the other hand, this is one of the regions where there is enhanced emission from hot molecular gas. Thus molecular gas seems to exist there, but is heated by local SN, as discussed in the previous paragraph. In summary, regarding the cold molecular gas distribution, while there is a similar distribution with the youngest star-forming regions globally around the ring, this does not seem to occur on scales of a few pcs.

The CNSFR is made up of small flocculent spirals or filament-like structures breaking into star-forming clumps extending along large sectors  ($\approx$ 300--400\,pc) of the ring (see Fig.{hst}). This has been interpreted as fragmentation of a gas disk due to gravitational instabilities \citep{colina00}. The fact that the range of estimated ages for the young star-forming regions in the south-west arc of the ring is very small, suggests that the fragmentation is taking place quasi-simultaneously (i.e. in less than few Myr) over sectors extending by hundred of parsecs. In addition, the east side of the ring appears to have formed stars some 10--15 Myr before the west side. Again the lack of an obvious age gradient, suggest that the star formation was essentially simultaneous over a large sector of the ring. In addition, if the ring is rotating around the nucleus as indicated by our stellar and gas kinematic maps -- which imply a rotational period of about 10\,Myr, a possible scenario is that the oldest stellar clusters were formed in the west side of the ring, but, after 10\,Myr have already almost completed one turn and are now observed in the south-east part of the ring.

\subsection{Mass of molecular and ionized gas in the ring} \label{AGN-SB}

The equations \ref{mhii} and \ref{mhm} can be used to calculate the total mass of ionized and hot molecular gas in the inner 9$^{\prime\prime}\times$9$^{\prime\prime}$ of NGC\,4303 covered by our field-of-view. By integrating the spectra over the total field of view and fitting the \hm\ and \br\ emission-line profiles from the resulting spectrum, we obtain $F_{H_{2}\lambda2.1218}\sim1.2\times10^{-14}$\,erg s$^{-1}$\,cm$^{-2}$ and   $F_{\rm Br\gamma}\sim8.5\times10^{-15}$\,erg s$^{-1}$\,cm$^{-2}$. The corresponding masses are M$_{\rm H_2}\sim110\,{\rm M_\odot}$ and M$_{\rm H II}\sim1.9\times10^{5}\,{\rm M_\odot}$. Both values are about three times the sum of the masses for the individual star-forming regions, shown in Table~\ref{tab-pars} and are in good agreement with the values found for the central region of active galaxies \citep[e.g.][]{n1068-exc,n5929}.

However, the M$_{\rm H_2}$ derived above may be just a small fraction of the total mass of molecular gas available in the central region of NGC\,4303 as the ratio between cold and hot molecular gas is usually in the range 10$^{5}-$10$^{7}$ \citep{dale05,ms06,mazzalay13}. Following  \citet{mazzalay13}, the mass of cold molecular gas can be obtained by:

\begin{equation}
\frac{M_{\rm cold}}{\rm M_{\odot}}\,\approx\,1174\times\left(\frac{L_{\rm {H_{2}}\,\lambda\,2.1218}}{L_{\odot}}\right),
\label{massafria}
\end{equation}
where $L_{\rm H_2}\,\lambda\,2.1218$ is the luminosity of the \hm\ line.  We derive $M_{\rm cold}\sim10^{8}\,{\rm M_{\odot}}$  and of $\sim$6.25 $\times$ 10$^6$ M$_{\odot}$ for the entire CNSFR and for the nucleus (60 pc radius), respectively. This large amount of molecular gas can be used to feed the AGN and/or to form stars. The estimated mass of cold H$_2$ gas for the entire CNSFR is in good agreement with the one obtained directly from CO observations of the nuclear disk of $6.9\times10^7$~M$_\odot$, as derived by \citet{schinnerer02}.

\subsection{Stellar and gas kinematics} \label{disc-stel}

\begin{figure*}
\centering
\includegraphics[scale=0.57]{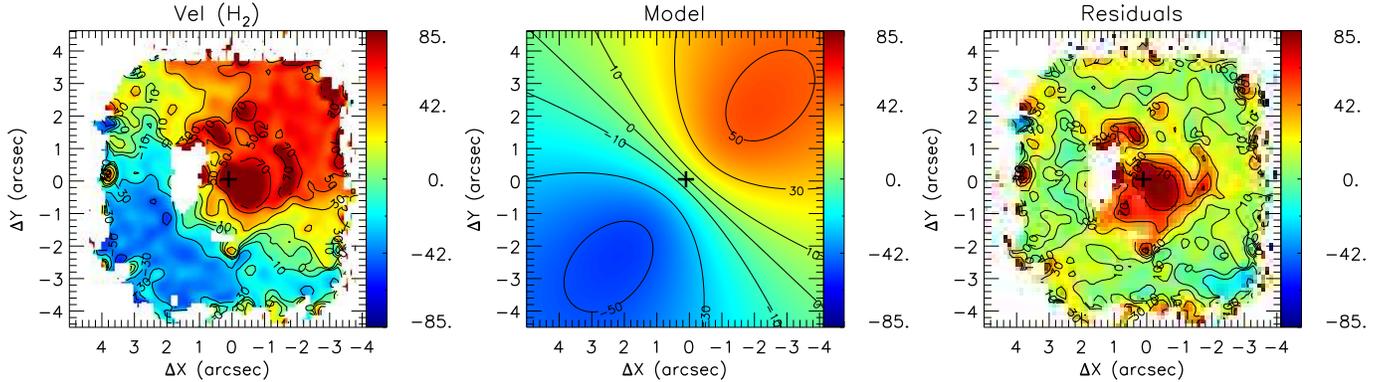}
\caption{Left: H$_2$ velocity field; middle: rotating disk model and right: residual map, obtained as the difference of the observed velocities and the model. The color bars show the velocity in units of \kms.} 
\label{resH2}
\end{figure*}

The stellar velocity field shown in the left panel of Fig.~\ref{stel} clearly presents a rotational component. In order to obtain relevant physical parameters, we fitted the stellar velocity field by a model of a thin disk, in which the stars have circular orbits in the plane of the galaxy, with velocities given by \citep{bertola91} 
\[  V(R,\Psi)=V_{s}+
\]
\begin{equation}
\frac{ARcos(\Psi - \Psi_0)sin(i) cos^p\theta}{{R^2[sin^2(\Psi - \Psi_0)+cos^2(i)cos^2(\Psi - \Psi_0)]+c^2_0cos^2(i)}^{p/2}}
\label{model-bertola}
\end{equation}
where $R$ and $\Psi$ are the coordinates of each pixel in the plane of the sky,  $A$ is the amplitude of the rotation curve, $\Psi_0$ is the position angle of the line of nodes, $V_s$ is the systemic velocity,  $i$ is the disk inclination relative to the plane of the sky, $c_0$ is a concentration parameter and $p$ is a model fitting parameter.

During the fit, the kinematical center was kept fixed at the location of the peak of the continuum emission, and the inclination of the disk was fixed to the inclination of the CNSFR $i=33^\circ$, obtained in Sec.~\ref{cnsfr-disc}.  
The difference between the observed and modeled velocities is smaller than 10\,\kms\ at most locations and thus we conclude that the adopted rotating disk model is a good representation of the observed velocity field. 
The best fitted model 
resulted in the following parameters: $V_s=1567\,\pm\,17\,$km\,$s^{-1}$, 
$\Psi_0=136^{\circ}\,\pm\,1^{\circ}$,
$A=160\,\pm\,7\,$km\,$s^{-1}$ and
$c_0=$2\farcs3$\,\pm\,0.2^{\prime\prime}$.  
Based on optical IFS with a coarser angular resolution and lower spectral resolution, \citet{colina99} showed that the gas velocity field of NGC\,4303 is consistent with a rotating disk with inclination of 45$^\circ$, distinct from the inclination of the large scale disk adopted here. Fixing the inclination of the disk to this value, we obtain similar physical parameters, except for the amplitude of the rotation curve that is smaller --  $A=132.1\,\pm\,5.2\,$km\,$s^{-1}$, but the quality of the fit is worse and thus we decided to adopt the inclination of the large scale disk.   
 
The stellar velocity dispersion map shows a ring of small values co-spatial with the ring of CNSFRs, showing that the stars still have the ''cold" kinematics of the raw gas that formed the stars. These low-$\sigma_*$ structures have been previously observed for Seyfert galaxies via similar near-IR integral field spectroscopy 
for which stellar population synthesis have confirmed that these structures are associated to intermediate-age (0.3-0.7 Gyr) stellar populations \citep[e.g.][]{mrk1066-pop}. In Dametto et al. (in preparation) we will present a detailed study of the optical/near-IR spectral energy distribution and verify the validity of this interpretation. At locations just beyond the CNSFR, the higher $\sigma_*$ values -- ranging from 70 to 100~\kms, can be attributed to the bulge of the galaxy.

The velocity fields for all emission lines are similar to that presented by \citet{colina99} on the base of optical observations, presenting a rotation pattern similar to that of the stars. We fitted the gas velocity field with the same model used to fit the stellar velocity field, keeping fixed all geometric parameters derived for the stars, allowing only the amplitude of the rotation curve to change. 
We confirm that this model reproduces well the various gas velocity fields, but with a larger velocity amplitude than that observed for the stars: $A=236\pm32$\kms\ for \hm\, $A=225\pm28$~\kms\ for \br\ and $A=151\pm25$~\kms\ for \feii\ (corrected for the inclination of the disk). Distinct rotational velocity amplitudes for the ionized and molecular gas as well as for the stars are commonly observed for active galaxies \citep[e.g.][]{mrk79,barbosa14}, likely indicating a slightly different three-dimensional distribution. 

The residuals between the measured and modeled gas velocities were usually smaller than $\approx$\,10--20\kms, except for the \hm\ velocity field shown in Fig.~\ref{resH2}. Redshift residuals of $\approx$80-100\kms\ are observed in a region extending to $\approx$\,120\,pc south-west of the nucleus, with similar orientation to that of the nuclear bar, revealing non-circular motions possibly associated with this bar. The velocity dispersion of \hm\ is also larger there, suggesting it may be due to the presence of more than one kinematic component (that our spectral resolution did not allow to separate). It is interesting to note that a high velocity dispersion was observed also in CO \citep{schinnerer02} in this region, that is identified with a possible molecular gas ``bridge" between the ring and the nucleus.

Regarding the ionized gas kinematics, \citet{colina99} has shown that the optical [O\,{\sc iii}] emission shows two components: one due to rotation in the  disk and another due to a compact outflow along the minor axis of the galaxy. Although we do not see an outflow in the gas velocity residuals, our data shows an increase in the \feii\ velocity dispersion to the east-northeast, where the \feii\ flux distribution shows an elongation at a similar orientation to that of the [O\,{\sc iii}] outflow.

Finally, although the stellar velocity dispersion shows lower velocity dispersion at the ring, the $\sigma$ maps for the gas (Fig.~\ref{gas-kin}) do not show a decrease neither an increase associated with the star-forming regions in the ring. This suggest that the star formation is not contributing significantly to the turbulence of the surrounding interstellar medium, as found also for a large sample of low-z LIRGs and ULIRGs \citep{arribas14} and in high-z galaxies \citep{genzel11}. The average velocity dispersion for the ionized gas (traced by the \br\ emission) of the CNSFR of NGC\,4303 is $\sigma=46\pm7$\,\kms, obtained by fitting the \br\ line profile integrated within a ring with inner radius of 1\farcs5 and outer radius of 4\farcs0, centred at the nucleus. This value is similar to that found for luminous star-forming clumps ($\sigma=43\pm3$\,\kms) of local (U)LIRGs \citep{arribas14}. 
If confirmed for larger samples of galaxies with CNSFR , it would indicate the velocity dispersion is dominated by global dynamical processes and less affected by local processes like star-formation on scales of few parsecs.

\section{Conclusions}

We have presented new emission-line flux and velocity maps, as well stellar velocity maps of the inner 0.7 kpc$\times$0.7 kpc of the nearby spiral galaxy NGC\,4303, at spatial resolutions of 78 pc, 47 pc, and 36 pc, for the J, H and K-bands, respectively, using near-IR integral field spectroscopy with the VLT instrument SINFONI. The observations cover the nucleus and circumnuclear star-forming ring (CNSFR) with radius $\approx$\,200--250\,pc. The main conclusions of this work are:

\begin{itemize}


\item The near-IR emission-line flux distributions delineate the CNSFR, with emission-line knots in the flux distributions of the different gas phases observed at different locations along the ring. The \hm\ and \feii\ emission lines show emission peaks at the nucleus, while the H\,{\sc i} recombination lines are dominated by emission from the CNSFR;

\item Along the ring, the strongest  H\,{\sc i} and \feii\ emission are observed mostly in the west side of the ring and at a region to the south-east, while the strongest emission in \hm\ seems to be anti-correlated with them, being observed mostly to the east;

\item The properties: \feii/\br\ and \hm/\br line ratios, EqW and $\sigma$ of the emission lines along the star-forming ring support an age difference between the west and east sides of the CNSFR, with the former being younger (2.5--7.5\,Myr) than the latter (10--25\,Myr);

\item The distribution of the star-forming regions and their age differences do not support fully the ``pop-corn'' and the ``pearls-in-a-string" scenarios for star formation in CNSFRs. The star formation in the CNSFR of NGC\,4303 appears to be instead episodic with stars forming quasi-simultaneously over a large sector of the ring (covering $\sim$\,300\,pc along the ring), aging as they rotate with an orbital time of several Myr;



\item Assuming the star-forming regions in the CNSFR are closer to instantaneous bursts with ages of about 4~Myr, we derive masses for the clusters in the range 0.3-1.5$ \times10^5$\,M$_\odot$. The corresponding masses for the  associated ionized and hot molecular gas are about $\sim0.25-1.2 \times 10^{4}$~M$_\odot$ and  $\sim0.25-0.5 \times 10^{1}$~M$_\odot$, respectively. For stellar populations with older ages of up to  10 Myr, the corresponding stellar masses will increase by up to factor of hundred; 

\item The \feii\ emission shows an elongation to $\approx$120\,pc north-east of the nucleus that could be associated with the previously known optical ([OIII]) outflow;

\item The \hm\ emission shows an elongation to $\approx$120\,pc west of the nucleus that could be the hot counterpart of the already known cold molecular gas ``bridge"  that connects the nucleus with the large circumnuclear molecular gas reservoir; 

\item The near-IR emission-line ratios (\feii/\br\ and \hm/\br) of the nucleus are consistent with the presence of an AGN and/or a SNe-dominated star-forming region. Since there is no evidence for an aged stellar cluster in the nucleus, the line ratios are interpreted as due to the combined effect of X-ray radiation and shocks at the base of the ionization cone of the AGN. Higher angular resolution spectroscopy is required to further explore this scenario;
      
\item The stellar velocity field is well reproduced by a model of a rotating disk with an inclination $i=33^\circ$ relative to the plane of the sky, major axis oriented along PA$\sim135^\circ$, and with a velocity amplitude of about 160\,\kms. The stars associated to the CNSFR show smaller velocity dispersion than the surroundings, revealing a cooler dynamical stellar population in the ring, consistent with their recent formation from cold gas;

\item The gas velocity fields are also dominated by rotation, similar to that observed for the stars but with a larger amplitude. A significant deviation from rotation was observed for the \hm\ emission in a region extending $\approx$\,120\,pc to the south-west, along the orientation of the nuclear bar. A higher \hm\ velocity dispersion is also observed at this location and is attributed to the presence of more than one kinematic component associated with non-circular motions along the nuclear bar.

\end{itemize}

All fits files for the emission-line flux distributions, velocity fields and velocity dispersion maps are available online as supplementary material.

\section*{Acknowledgments}
We thank an anonymous referee for useful suggestions which helped to improve the paper. 
R.A.R. acknowledges support from FAPERGS (project N0. 2366-2551/14-0) and CNPq (project N0. 470090/2013-8 and 302683/2013-5). 
L.C. acknowledges support from CNPq special visitor fellowship  PVE  313945/2013-6
 under  the  Brazilian  program  Science  without Borders.  L.C.,  J.P.,  and  S.A.
are  supported  by  grants  AYA2012-32295,  AYA2012-39408 and ESP2015-68964 from the Ministerio de Econom\'ia
y Competitividad of Spain.
D.A.S., R.R. and T.S.B. thank to CNPq for financial support.

\label{lastpage}

\end{document}